\documentclass[prb, twocolumn, nofootinbib]{revtex4-2}

\usepackage{bibunits}
\defaultbibliographystyle{apsrev4-2}
\renewcommand{\citenum}[1]{\onlinecite{#1}}

\usepackage{amssymb}
\usepackage{amsmath}
\usepackage{bbold}
\usepackage{mathrsfs} 
\usepackage{graphicx}
\usepackage{bm}
\usepackage[x11names]{xcolor}
\usepackage[unicode=true,colorlinks=true]{hyperref} 
\hypersetup{urlcolor = blue, linkcolor=blue }

\newcommand{\KP}{\texorpdfstring{{\bf k}\ensuremath{\cdot}{\bf p}}{k.p}}

\begin{document}

\title{Tailoring the electron and hole Land{\'e} factors in lead halide perovskite nanocrystals by quantum confinement and halide exchange }
\makeatletter \let\thetitle\@title \makeatother

\author{Mikhail~O.~Nestoklon$^{1}$}
\email{nestoklon@gmail.com}
\author{Erik~Kirstein$^1$}
\author{Dmitri~R.~Yakovlev$^{1,2}$}
\email{dmitri.yakovlev@tu-dortmund.de}
\author{Evgeny~A.~Zhukov$^{1}$}
\author{Mikhail~M.~Glazov$^2$}
\author{Marina~A.~Semina$^2$}
\author{Eugeniyus~L.~Ivchenko$^2$}
\author{Elena~V.~Kolobkova$^{3,4}$}
\author{Maria~S.~Kuznetsova$^5$}
\author{Manfred~Bayer$^{1}$}
\affiliation{
  $^1$~Experimentelle Physik 2, Technische Universit{\"a}t Dortmund, 44227 Dortmund, Germany\\
  $^2$~Ioffe Institute, Russian Academy of Sciences, 194021 St. Petersburg, Russia\\
  $^{3}$ITMO University, 199034 St. Petersburg, Russia\\
  $^{4}$St. Petersburg State Institute of Technology, 190013 St. Petersburg, Russia\\
  $^{5}$Spin Optics Laboratory, St. Petersburg State University, 198504 St. Petersburg, Russia  
}

\makeatletter 
    
\renewcommand\onecolumngrid{
\do@columngrid{one}{\@ne}%
\def\set@footnotewidth{\onecolumngrid}
\def\footnoterule{\kern-6pt\hrule width 1.5in\kern6pt}%
}

\renewcommand\twocolumngrid{
        \def\footnoterule{
        \dimen@\skip\footins\divide\dimen@\thr@@
        \kern-\dimen@\hrule width.5in\kern\dimen@}
        \do@columngrid{mlt}{\tw@}
}%

\makeatother    




\begin{abstract} 
The tunability of the optical properties of lead halide perovskite nanocrystals makes them highly appealing for applications. Both, halide anion exchange and quantum confinement pave the way for tailoring their band gap energy. For spintronics applications, the Land{\'e} $g$-factors of electrons and hole are of great importance. By means of the empirical tight-binding and \KP\ methods, we calculate them for nanocrystals of the class of all-inorganic lead halide perovskites CsPb$X_3$ ($X=$ I, Br, Cl). The hole $g$-factor as function of the band gap follows the universal dependence found for bulk perovskites, while for the electrons a considerable modification is predicted. Based on the \KP\ analysis we conclude that this difference arises from the interaction of the bottom conduction band with the spin-orbit split electron states. The model predictions are confirmed by experimental data for the electron and hole $g$-factors in CsPbI$_3$ nanocrystals placed in a glass matrix, measured by time-resolved Faraday ellipticity in a magnetic field at cryogenic temperatures. 
\end{abstract}

\maketitle
\begin{bibunit}
\section{Introduction}  

The development of quantum technologies and spintronics calls for a search of suitable material platforms. The recently emerging lead halide perovskite semiconductors~\cite{Vardeny2022_book,Vinattieri2021_book} offer a new testbed for spin-dependent phenomena. Among them, the optical orientation of charge carrier spins, their coherent spin precession, spin mode-locking, spin-flip Raman scattering, and dynamic nuclear polarization have been demonstrated for bulk lead halide perovskites~\cite{Giovanni2015,zhang2015,Wang2018,Odenthal2017,Belykh19,kirstein2022am,Kirstein22} and their nanocrystals~\cite{Nestoklon18,strohmair2020,Crane20,Grigoryev21,Kirstein22_ml}. Information on the electron and hole Land{\'e} factors ($g$-factors), which determine the Zeeman splitting of their spin states in an external magnetic field, is of key importance for understanding the spin-dependent phenomena and for resulting spintronic applications. The magnetic field allows one to unravel otherwise hidden information on the band structure, \textit{e.g.}, by splitting degenerate spin states and activating optically-forbidden dark exciton states~\cite{Tamarat20}. Electron and hole $g$-factors were measured for a set of bulk hybrid organic-inorganic and all-inorganic lead halide perovskites by pump-probe Kerr rotation and spin-flip Raman scattering~\cite{Kirstein22}. It was shown that the carrier $g$-factors obey a universal dependence on the band gap energy across different perovskite material classes, which can be summarized in a universal semi-phenomenological expression. This empirical result was corroborated by atomistic calculations based on the combination of the density functional theory (DFT) and empirical tight-binding (ETB) methods.

Lead halide perovskite nanocrystals (NCs) have attracted a lot of attention during the last decade due to their excellent optical properties and technologically simple synthesis \cite{Protesescu15,Yakunin15,Dey21,Younis21}. The quantum confinement in NCs strongly affects the electronic and optical properties. It is known, that in conventional semiconductor quantum dots the confinement leads to a renormalization of the carriers' $g$-factors \cite{ivchenko05a}. This effect has been studied in nanostructures of III--V and II--VI semiconductors and is well-understood from the \KP\ theory \cite{Kiselev98} and atomistic calculations \cite{Tadjine17}. Universal dependences of the $g$-factors on the band gap have been demonstrated for quantum wells \cite{Yugova07,Sirenko1997} and III--V quantum dots~\cite{Tadjine17}, however, in a much narrower range of band gap variations compared to bulk perovskites~\cite{Kirstein22}. All that calls for an investigation of the $g$-factors in perovskite NCs.

Modern calculation methods allow one to compute the electronic band structure of bulk semiconductors with reasonable precision \cite{Heyd06}, but the use of DFT for nanostructures is challenging. For a qualitative analysis of nanostructures with sizes of tens of nanometres, the effective mass approximation~\cite{Luttinger55} is the method of choice, while for nanostructures of few nm size empirical atomistic methods are preferable~\cite{Zunger_PP,Xavier_book}. The empirical tight-binding method is one of the simplest approximations suitable for an accurate description of the band structure of conventional semiconductors at the lowest possible computation cost~\cite{Xavier_book}. The ETB method within the $sp^3d^5s^*$ basis in the nearest neighbor approximation gives a precise description of the band structure of bulk III--V~\cite{Jancu98} and group IV~\cite{Niquet09} semiconductors. It has been shown recently that this method can be used to model the band structure of inorganic lead halide perovskites with meV-range precision~\cite{Nestoklon21}.

In this study we use the ETB method within the $sp^3d^5s^*$ basis applying the nearest neighbor approximation to calculate the electron and hole $g$-factors in all-inorganic lead halide perovskite NCs based on CsPbI$_3$, CsPbBr$_3$ and CsPbCl$_3$. We show that the hole $g$-factors in NCs follow the empirical trend, which was first observed for bulk perovskites\cite{Kirstein22}. The electron $g$-factors show a significant renormalization, which results in a deviation from the bulk empirical trend. Using the \KP\ calculations we uncover the origin of this behavior and show that it is due the quantum-confinement induced admixing of the excited heavy- and light-electron states to the ground electron state. We measured the electron and hole $g$-factors by means of time-resolved Faraday ellipticity on ensembles of CsPbI$_3$ NCs of different sizes. The measured $g$-factors are in good agreement with the theory predictions. 


\section{Empirical tight-binding calculations}
\label{seq:ETB_NCs} 

We use an \textit{ab initio} inspired ETB method to compute the band gap and $g$-factors of charge carriers quantum-confined in cubic-shaped nanocrystals of CsPb$X$\textsubscript{3} ($X=$ I, Br, Cl) in the cubic phase. In the ETB method,\cite{Lowdin50} the $i$-th electron wave function $\Phi_i ({\bf r})$
is expanded in the basis of orthogonal atomic-like functions $\phi_{\alpha}({\bf r})$ :
\begin{equation}\label{eq:TB_psi_r}
  \Phi_i ({\bf r}) = \sum_{n \alpha} C_{n \alpha}^i \phi_{\alpha}({\bf r} - {\bf r}_{n} )\:,
\end{equation}
where $n$ enumerates the atoms with coordinates ${\bf r}_{n}$, the index $\alpha$ runs through different orbitals and spins, and $C_{n \alpha}^i$ are expansion coefficients. We use the $sp^3d^5s^*$ variant of the method \cite{Jancu98} with twenty orbitals per atom. As demonstrated in Refs.~\citenum{Nestoklon21,Kirstein22}, this method provides an accurate description of the band structure of lead halide perovskites. In Ref.~\citenum{Nestoklon21} it is also demonstrated that the method gives a reasonable description of the Pb-plane terminated surface without considering additional passivation and reconstruction.  


We start from the calculation of the electron and hole states in NCs. In the basis \eqref{eq:TB_psi_r}, the Schr\"odinger equation for the nanocrystal reduces to the eigenvalue problem for a sparse matrix:
\begin{equation}\label{eq:TB_Ham_r} 
  \sum_{n',\varsigma} H_{n\alpha,n'\varsigma} C^i_{n'\varsigma} = E_i C^i_{n\alpha} \,.
\end{equation}
Here $H_{n\alpha,n'\varsigma}$ are the matrix elements of the ETB Hamiltonian calculated following Ref.~\citenum{Slater54} and $E_i$ is the energy of the $i$-th electron state. The solutions of Eq.~\eqref{eq:TB_Ham_r} near the band-edge can be found using the Thick-Restart Lanczos algorithm \cite{Wu00}.

\begin{figure*}
  \centering{\includegraphics[width=1\linewidth]{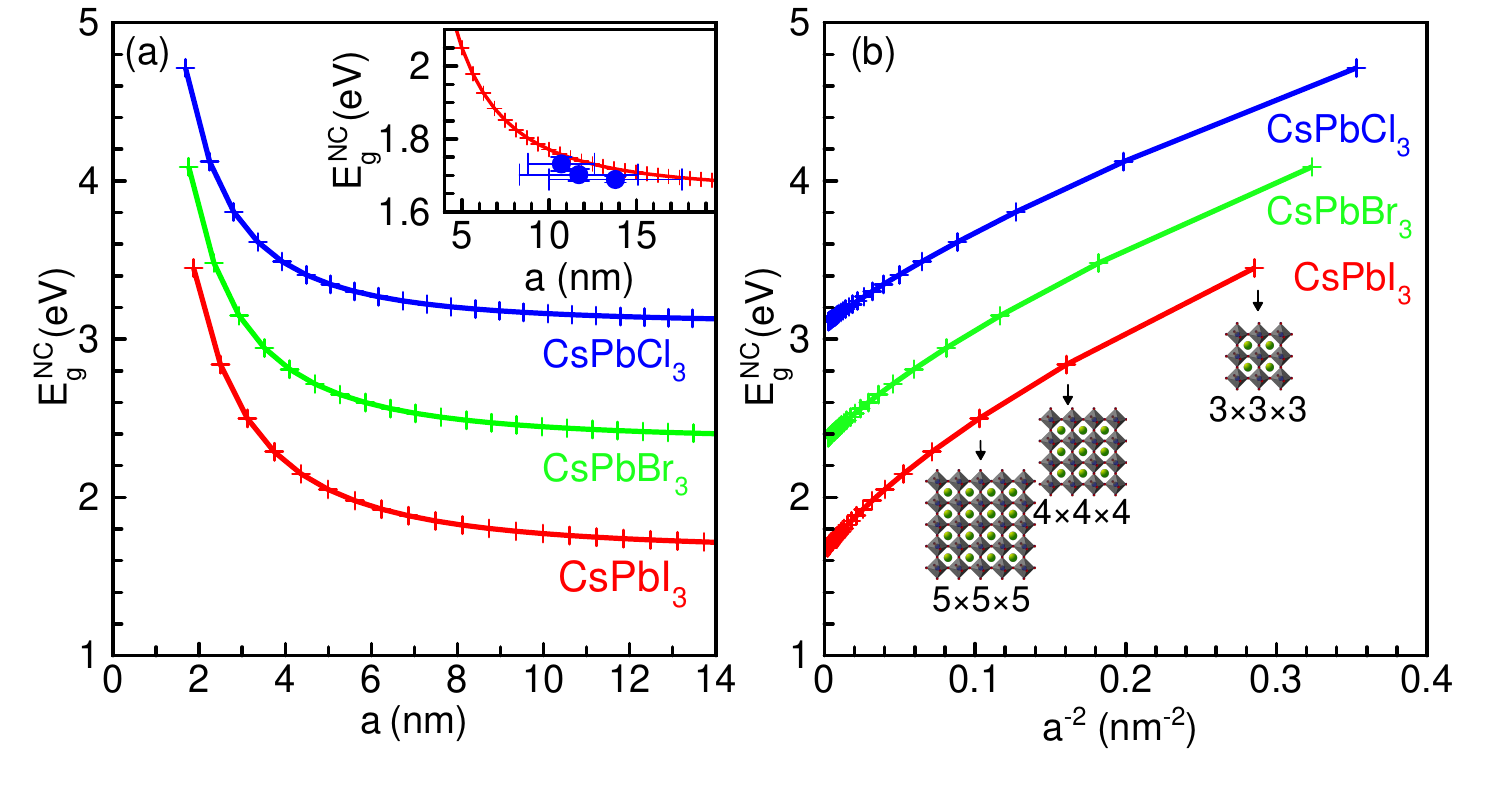}}
 \caption{
Calculated band gap of NCs as a function of NC edge length $a$ (panel (a)) and of $1/a^2$ (panel (b)). In the ETB calculations the integer number of monolayers is multiplied with the lattice constant $a_0 = 0.561$\,nm, $0.586$\,nm, $0.624$\,nm for CsPb$X_3$ with $X=$ Cl, Br and I, respectively~\cite{Becker18}. The crosses show the results of numerical calculations and the lines are guides to the eye. In the inset, a comparison of the calculated band gap energies (red line) with experimental data for the CsPbI$_3$ NCs samples used in this study (blue dots) is shown. Details of the sample characterization are given in the Supporting Information S4, the NC size was evaluated via STEM, and the energy is given by the maximum in Faraday ellipticity spectra of Figure~\ref{fig:experiment}a. The error bars give the half width at half maximum of both the Faraday ellipticity spectra (Figure~\ref{fig:experiment}a) and the STEM characterization (Supporting Information Figure~S6).
		}
\label{fig:Eg}
\end{figure*}

The effective band gap $E_g^{\rm NC}=E_g+E_{\rm e}+E_{\rm h}$ is the sum of the bulk material band gap $E_g$ and the electron and hole confinement energies, $E_{\rm e}$ and $E_{\rm h}$, respectively. It is extracted as the energy difference between the lowest confined conduction band state and the uppermost confined valence band state, and depends on the NC size according to an almost linear function of $1/a^2$, where $a$ is the edge length of the cubic NC, see Figure~\ref{fig:Eg}. This is in line with the effective mass approximation and with the \KP\ calculations of CsPbI$_3$ nanocrystals in Ref.~\citenum{Zhao20}. For all three materials the confinement-induced band gap renormalization is significant and reaches $\sim 1$\,eV for small nanocrystal sizes with $a\sim 3$\,nm.
In the inset of Figure~\ref{fig:Eg}a we compare the calculated band gap energy as function of the NC size with the corresponding dependence of the maximum energy of the Faraday ellipticity amplitude, measured at low temperature (see below). The size of the NCs is evaluated via scanning transmission electron microscopy (STEM) at room temperature, see Supplementary Information S4. The agreement between calculation and experiment is good, the difference ($\sim30$~meV) may be attributed to the neglected exciton binding energy, the effect of the dielectric contrast not included in the modeling, or underestimated masses of the carriers in our ETB parametrization. However, the determination of the experimental band gap energy is also an involved problem as the photoluminescence line is much broader than the Faraday ellipticity spectrum for the measured NCs.

The electron and hole $g$-factors are calculated by introducing a weak magnetic field in the ETB Hamiltonian \eqref{eq:TB_Ham_r} following a standard procedure by using the Peierls substitution\cite{Graf95,Kim21}: the vector-potential-dependent phase, and the Zeeman term, $\mu_B(\bm{\sigma}\cdot{\bf B})$, where $\mu_B$ is the Bohr magneton, $\bm{\sigma}$ is the vector of Pauli matrices describing the spin and ${\bf B}$ is the magnetic field, are added to the off-diagonal and diagonal matrix elements of the tight-binding Hamiltonian~\eqref{eq:TB_Ham_r}, respectively. 

For bulk, the $g$-factors of different materials show trends which may be understood in the framework of \KP\ theory.\cite{Kirstein22} The value of the bulk electron $g$-factors is mostly determined by a constant contribution (given by the bare $g$-factor and the interaction with remote bands) and the interaction with the valence band which changes with the band gap. The value of the hole $g$-factors is given by the constant bare $g$-factor and the interaction with the three low-lying conduction bands, the latter contribution is band-gap-dependent. The similar consideration may be applied for III-V semiconductors and used to account for the change of $g$-factors in nanostructures due to quantum confinement. The goal of our study is to accurately check whether such generalization may be done also for perovskite nanostructures.

\begin{figure*}
  \centering{\includegraphics[width=0.9\linewidth]{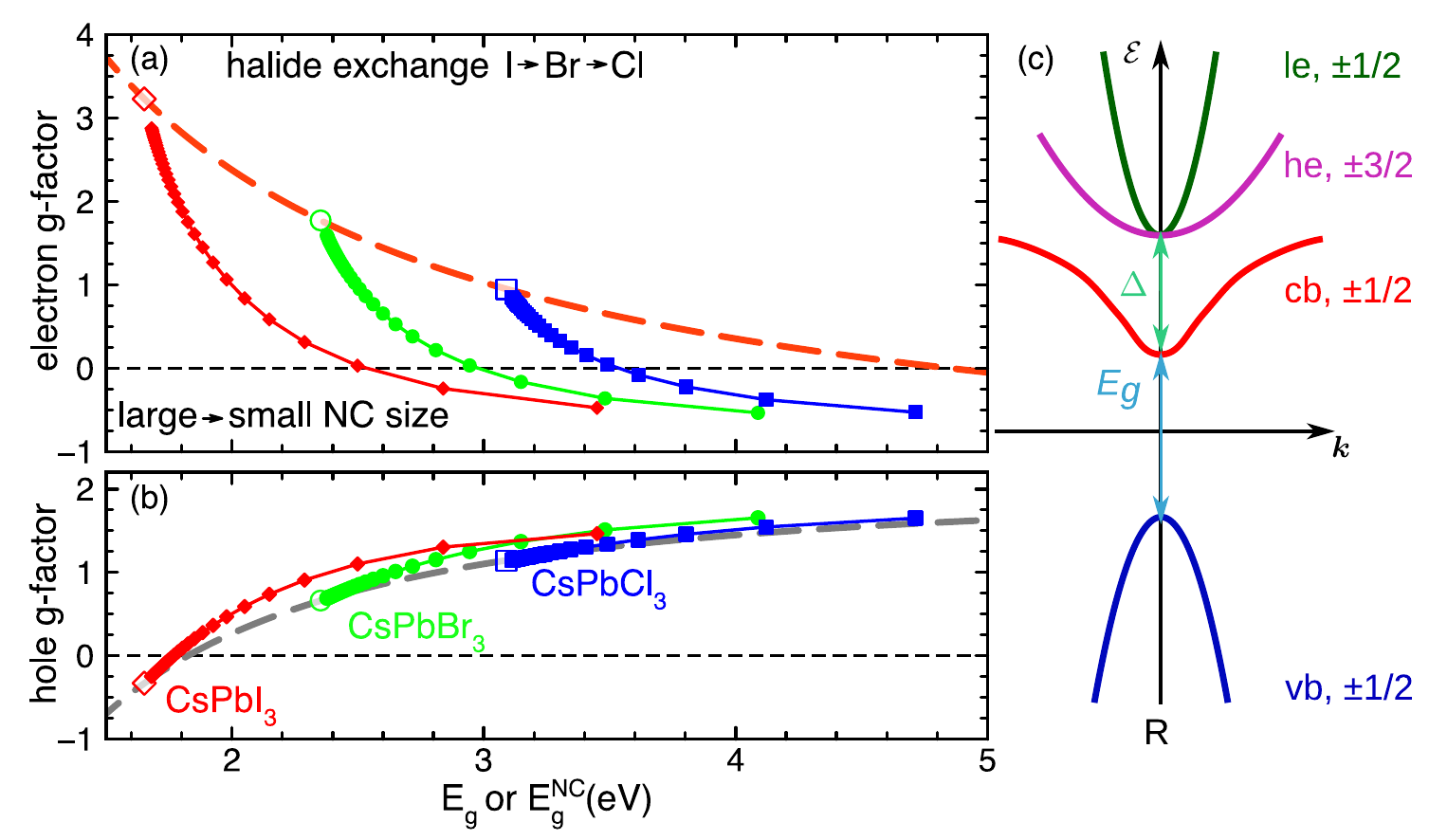}}
  \caption{ 
The closed symbols show the $g$-factors of electrons (a) and holes (b), calculated for various pervoskite NCs by ETB. The open symbols show the $g$-factors calculated in ETB for the bulk crystals. The orange and grey dashed lines are the results of a \KP\ model fit to reproduce the experimental data for bulk crystals, see Ref.~\citenum{Kirstein22}. (c) Sketch of the band structure of cubic lead halide perovskites in vicinity of the direct band gap at the $R$ point. 
  }\label{fig:g_vs_Eg}
\end{figure*}

The resulting values of the $g$-factors of the electron and hole ground states as function of the effective band gap in NCs are presented in Figure~\ref{fig:g_vs_Eg} by symbols. For a given material, with increasing $E_g$ the electron $g$-factor decreases whereas the hole $g$ factor increases. The $g$-factors are isotropic, which is expected for cubic nanocrystals in the cubic phase. Note that we use the updated ETB parametrization which better reproduces the $g$-factors of bulk perovskites, see \nameref{sec:meth} for details.

In bulk lead halide perovskites, a universal dependence of the charge carrier $g$-factors on the band gap has been found~\cite{Kirstein22}. The hole $g$-factor in NCs only weakly deviates from this universal curve shown by the dashed line in Figure~\ref{fig:g_vs_Eg}b. On the other hand, the electron $g$-factor shows a strong effect of the quantum confinement, dropping significantly with an increase of the effective band gap (a decrease of the NC size) and even changing sign from positive to negative, see Figure~\ref{fig:g_vs_Eg}a.

To provide a qualitative interpretation of the results of the ETB calculations and reveal the origin of the deviation from the universal dependence, we extend the \KP\ model\cite{Kirstein22} to account for the size quantization in perovskite NCs. For the hole $g$-factors, straightforward application of the Kane model for spherical NCs with infinite barriers yields (see Supporting Information S2 and Refs.~\citenum{Kiselev98,ivchenko05a} for details)
\begin{equation}
\label{eq:gh}
g_{\rm h}(E_{\rm h}) = 2 - \frac{4}{3} \frac{p^2}{m_0}w_{\rm h}\biggl(\underbrace{\frac{1}{E_g+E_{\rm h}}}_{vb-cb}-\underbrace{\frac{1}{E_g+E_{\rm h}+\Delta}}_{vb-(he/le)}\biggr)\,.
\end{equation}
Here $p$ is the interband momentum matrix element, $m_0$ is the free-electron mass, $E_g$ and $\Delta$ are the band gap and  spin-orbit splitting in the bulk material, respectively, and $E_{\rm h}$ is the hole size quantization energy. The factor $w_{\rm h}= \int d^3r f_{\rm h}^2(r)\leqslant 1$, where $f_{\rm h}(r)$ is the valence band envelope function, accounts for the confinement-induced band mixing. Both energy denominators and $w_{\rm h}$ depend on the hole size-quantization energy.   

The significant renormalization of the conduction band $g$-factors results from the complex conduction band structure: In lead halide perovskites the lowest conduction band is formed by spin 1/2 states and the next bands represent heavy and light electrons with spin $3/2$. The magneto-induced mixing with the spin-orbit split-off heavy- and light-electron bands results in the electron $g$-factor renormalization. The calculation within the \KP\ model, which explicitly accounts for the band mixing (Supporting Information S2), results in the following expression for the electron $g$-factor
\begin{equation}
\label{eq:ge}
g_{\rm e}(E_{\rm e}) = -\frac{2}{3} + \frac{4}{3} \frac{p^2}{m_0}\underbrace{\frac{w_{\rm e}}{E_g+E_{\rm e}}}_{cb-vb} + \Delta g_{\rm remote} + \delta g_{\rm e}^{\rm so},
\end{equation}
with $w_{\rm e}$ accounting for the confinement-induced band mixing, $\Delta g_{\rm remote}$ being the remote bands' contribution~\cite{Kirstein22}, and $\delta g^{\rm so}_{\rm e}$ being the contribution which arises from the size-quantization induced mixing with the split-off electron band 
\begin{equation}
\label{genc:text}
\delta g_{\rm e}^{\rm so} =  -40 \frac{\bar\gamma^2}{\gamma_1} \frac{E_{\rm e}}{\Delta} \, ,
\end{equation}
$\gamma_1$ and $\bar\gamma$ are the Luttinger parameters. While deriving
Eq.~\eqref{genc:text} we assumed that $E_{\rm e}/\Delta \ll 1$ (see Supporting Information S2 for the complete analysis and comparison with ETB). Equations~\eqref{eq:ge} and \eqref{genc:text} explain the strong renormalization of the electron $g$-factor provided by the quantum confinement: the $g$-factor significantly decreases with increasing electron size-quantization energy, in agreement with the ETB calculations.

\section{$g$-factors measured in \NoCaseChange{CsPbI}\texorpdfstring{\textsubscript{3}}{3} \NoCaseChange{NCs}}

Published  information on the electron and hole $g$-factors in lead halide perovskite NCs is limited. The reported experimental data are related to specific sizes of CsPbBr$_3$ and CsPb(Br,Cl)$_3$ NCs~\cite{fu2017,Isarov2017,Crane20,Grigoryev21,Kirstein22_ml}, but no systematic study on their size and band gap dependence is available. In order to check the role of quantum confinement for the carrier $g$-factors that we predict by model calculations, we experimentally examine a set of CsPbI$_3$ nanocrystals embedded in a fluorophosphate glass matrix. Three samples with NC sizes varying in the range of  8--16~nm are studied. Their NC size distributions are centered at 13.8~nm (sample \#1), 11.7~nm (sample \#2), and 10.7~nm (sample \#3), see Supporting Information S4. The NC size was changed by the synthesis procedure (\nameref{sec:meth}), providing a rather broad size dispersion within each sample. By tuning the laser photon energy we selectively address NCs with specific mean sizes directly related to the exciton transition energy, Figure~\ref{fig:Eg}.
 
\begin{figure*}%
\includegraphics[width=0.8\textwidth]{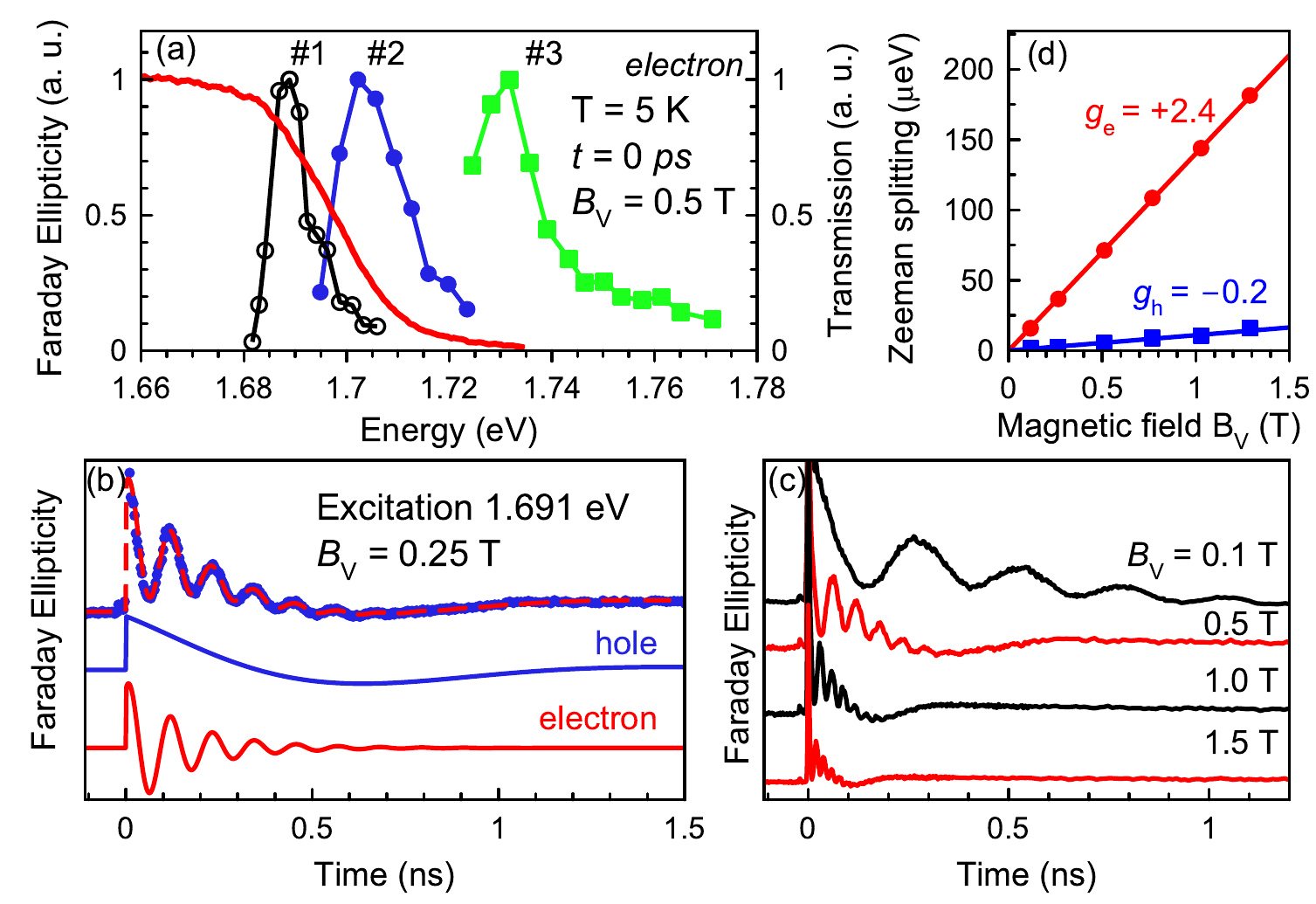}
\caption{(a) Transmission spectrum of CsPbI$_3$ NCs (sample \#1, red line, right axis) and spectral dependences of Faraday ellipticity amplitudes of the electron signal measured for the three studied samples (for $B_{\rm V}=0.5$\,T at zero time delay, left axis). $T=5$\,K.
(b)~Faraday ellipticity dynamics (blue dots) measured for sample \#1 at 1.691\,eV in $B_{\rm V}=0.25$\,T. The red dashed line is a fit with two components using the equation for the signal $A_{\rm FE}$ given in \nameref{sec:meth}. The individual hole and electron components are given below. 
(c) Faraday ellipticity dynamics measured for sample \#1 at different magnetic fields.
(d) Magnetic field dependence of the electron and hole Zeeman splittings. Experimental data given by the symbols and lines are linear fits to evaluate the corresponding $g$-factors. The sign of the hole $g$-factor is determined from another experiment. 
}
\label{fig:experiment}%
\end{figure*}

In Figure~\ref{fig:experiment}a a transmission spectrum measured for sample \#1 at  $T=5$\,K is shown. Its edge, centered at 1.697\,eV, originates from the exciton absorption. The spread of the absorption edge by about 26\,meV from 1.684 to 1.710\,eV is due to the NC size dispersion. 

We apply the time-resolved Faraday ellipticity (TRFE) technique to measure the electron and hole $g$-factors (\nameref{sec:meth}). This is an all-optical pump-probe technique using pulsed lasers to address the coherent spin dynamics of carriers in a magnetic field~\cite{Yakovlev_Ch6}. Recently, this technique was successfully used for investigation of CsPb(Cl,Br)$_3$ NCs in glass~\cite{Kirstein22_ml} and CsPbBr$_3$ solution-grown NCs~\cite{Crane20,Grigoryev21}. Thereby, spin-oriented carriers are photogenerated by circularly-polarized pump pulses and the dynamics of their spin polarization are detected through variations of the ellipticity of the linearly polarized probe pulses~\cite{YugovaPRB09}. Spectrally narrow (1.5\,meV width) laser pulses with duration of 1.5\,ps (repetition rate of 76\,MHz) were used. The laser photon energy was scanned across the exciton absorption band, exciting narrow subsets of NC sizes, so that we could measure the spectral dependences of the $g$-factors and compare with the modeling results for their dependence on $E_g^{\rm NC}$, i.e. on the carrier confinement energies.

A typical TRFE dynamics trace measured on sample \#1 at the laser photon energy of 1.691\,eV is shown in Figure~\ref{fig:experiment}b. The measurement is performed in a magnetic field of $B_{\rm V}=0.25$\,T applied perpendicular to the light wavevector (Voigt geometry). The carrier spins, which are initially oriented along the wavevector, undergo Larmor precession about the magnetic field. In the TRFE dynamics this leads to oscillating signals, which are decaying with the spin dephasing time $T^*_2$. As is typical for bulk lead halide perovskites~\cite{kirstein2022am} and also their NCs~\cite{Grigoryev21}, the spin dynamics are contributed by two oscillating signals provided by the electrons and the holes. We fit the experimental dynamics (blue circles in Figure~\ref{fig:experiment}b) by a function accounting for these two decaying oscillatory components (\nameref{sec:meth}). The fit is shown by the dashed red line, and the individual electron and hole dynamics are given by the solid lines in Figure~\ref{fig:experiment}b. The following parameters are evaluated from the fit: for the electrons, the Larmor precession frequency $\omega_{\rm L, e}=55.9$\,rad/ns and $T_{\rm 2,e}^*=170$\,ps, and for the holes, $\omega_{\rm L, h}=3.9$\,rad/ns and $T_{\rm 2,h}^*=500$\,ps. Their amplitudes are about equal: $S_{\rm e} \approx S_{\rm h}$. The electron $g$-factor is always positive for these energies~\cite{Kirstein22}. 

With increasing magnetic field the Larmor precession frequencies increase and the spin dephasing times shorten, as one can see in Figure~\ref{fig:experiment}c. The magnetic field dependences of the electron and hole Zeeman splittings evaluated by using $E_{\rm Z, e(h)} = \hbar \omega_{\rm L, e(h)}$ are shown in Figure~\ref{fig:experiment}d. Both dependences can be fitted with a linear function, which allows us to evaluate the $g$-factors according to $g_{\rm e(h)} = E_{\rm Z, e(h)} / ( \mu_{\rm B} B_{\rm V})$:  $g_{\rm e}=+2.4$ and $g_{\rm h}=-0.2$. We assign the fast Larmor precession frequency to the electrons and the slow one to the holes, referring to the model predictions (Figure~\ref{fig:g_vs_Eg}) and also comparing with the experimental data for bulk materials having comparable band gaps~\cite{Kirstein22}. Note that the TRFE is insensitive to the $g$-factor sign, but the latter can be obtained from the spectral dependence of the $g$-factor value and the model considerations. Namely, for $g_{\rm h}<0$ its absolute value decreases with growing $E_g^{\rm NC}$, which is the case for the studied CsPbI$_3$ NCs, but for $g_{\rm h}>0$ it increases, see Figure~\ref{fig:g_vs_Eg} and Ref.~\citenum{Kirstein22} for details. 

Very importantly, the dependences $E_{\rm Z, e(h)}(B_{\rm V})$ shown in Figure~\ref{fig:experiment}d has no offset for extrapolation to zero magnetic field. This allows us to assign the measured spin dynamics to resident electrons and holes confined in the NCs and not to carriers bound in excitons. In the latter case the electron-hole exchange interaction, which in perovskite NCs amounts to $0.5-2.0$\,meV~\cite{Tamarat20}, would result in spin beats with a single-exciton frequency. 
We suggest that the resident carriers in the NCs appear from long-living photocharging, where either the electron or the hole from a photogenerated electron-hole pair escapes from the NC core. As a result, some NCs in the ensemble become charged with electrons, some with holes, while the rest stays neutral. This situation has been studied for CsPbBr$_3$ NCs synthesized in solution and more arguments for its validity can be found in Ref.~\citenum{Grigoryev21}. 

\begin{figure}%
\includegraphics[width=\linewidth]{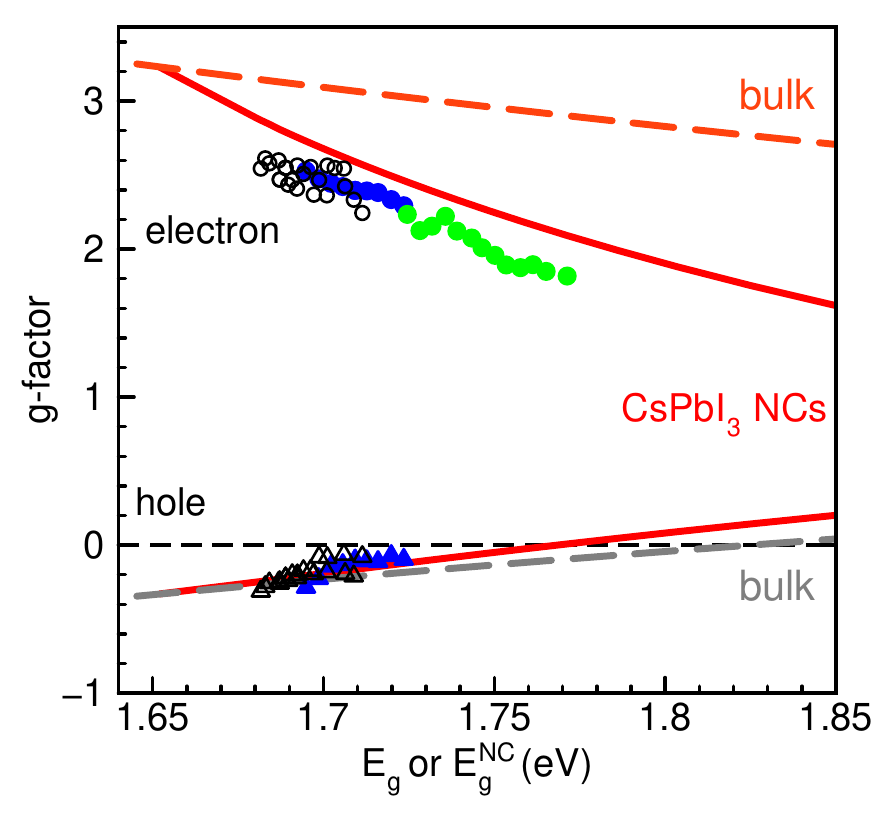}
\caption{Energy dependence of the electron (circles) and hole (triangles) $g$-factors measured at $T=5$\,K in CsPbI$_3$ NCs: data from samples \#1 (black open  symbols), \#2 (blue symbols) and \#3 (green symbols). Red lines are ETB model results for NCs. The orange and grey dashed lines are the universal dependences for bulk crystals taken from Ref.~\citenum{Kirstein22}.
}
\label{fig:factors_experiment}%
\end{figure}

We measured three samples of CsPbI$_3$ NCs by TRFE and tuned the laser photon energy across the exciton absorption band of the NC ensemble. The spectral dependences of the electron spin amplitudes are shown in Figure~\ref{fig:experiment}a. The amplitudes are normalized on their largest value for each sample. The respective maxima are located at 1.690\,eV (sample \#1), 1.705\,eV (sample \#2), and 1.735\,eV (sample \#3). The spin dynamics  were detected in the spectral range from 1.682 to 1.772\,eV corresponding to NC sizes of $8-16$~nm, see the insert of Figure~\ref{fig:Eg}a. The measured electron and hole $g$-factors are plotted in Figure~\ref{fig:factors_experiment}. Note, that the hole signal was not detectable in the sample \#3 so that we show only data for its electron $g$-factor.

In order to compare the experimental data with the theoretical predictions, we show in Figure~\ref{fig:factors_experiment} the model calculations taken from Figure~\ref{fig:g_vs_Eg}. One can see that in agreement with our ETB calculations and the qualitative \KP\ analysis, the experimental data for CsPbI$_3$ NCs clearly show a strong deviation of the electron $g$-factor from the universal bulk dependence (orange dashed line), while the hole $g$-factor closely follows this dependence (grey dashed line). Note that the clear deviation from the univeral behaviour is seen even for relatively large nanocrystals and can be naturally explained by admixture of the split-off bands to the ground electron band by the size quantization (Supporting Information S2).

Care has to be exercised when comparing the experimental data with the calculations. In Figure~\ref{fig:factors_experiment} we show the experimental data as a function of excitation energy, corresponding to the ground state energy of the excitons in NCs, while the results of the calculations are shown as function of the effective band gap of the NCs $E_g^{\rm NC}$. We argue that the exciton binding energy does not change the results and their interpretation qualitatively as the binding energy is not too large (of the order of a few tens of meV) and leads to an overall shift, which smoothly depends on NC size. 

\section{Discussion}

The obtained theoretical results match well the experimental findings. Previously, it had been reported that the experimentally observed\cite{Crane20,Grigoryev21} electron $g$-factors in large CsPbBr$_3$ NCs are significantly ($\sim 10\%$) lower than the bulk value\cite{Belykh19} while the hole $g$-factor is marginally larger. Our results provide the explanation for this observation and demonstrate that this is a general feature common for perovskite-based NCs, which originates from the band structure of these materials. Based on the ETB and \KP\ calculations we have demonstrated that the significant deviation of the electron $g$-factor from the universal bulk dependence~\cite{Kirstein22} arises from the quantum-confinement induced admixture of the heavy- and light-electron bands to the electron ground state.

The calculations were performed for cubic shaped NCs, while for NCs in glass, due to the growth method, droplets of pseudo spherical shape are expected as confirmed by STEM, see Supplementary Information S4. Still, the ETB calculations agree well with the experimental data and we expect that the $g$-factor does not significantly depend on NC shape.\cite{Avdeev23} The elongated shape and low-symmetry crystal phase\cite{Kirstein22,BenAich2020} of NCs both lead to an anisotropy of the $g$-factor, these effects and their interdependences are yet to be investigated. However, the main trends demonstrated here are not affected by this anisotropy. Nevertheless, the random orientation of the NC anisotropy axes in an ensemble should be accounted in the analysis of the experimental data as it might lead to small uncertainties of the measured $g$-factor value presented in Figure~\ref{fig:factors_experiment}. 

\section{Conclusions} 

In conclusion, we have analyzed theoretically the role of confinement and halide exchange on the electron and hole $g$-factors in lead halide perovskite nanocrystals. Both the empirical tight-binding and \KP\ approaches suggest a strong deviation of the electron $g$-factor from the bulk values of the previously established universal dependence of the $g$-factor on the band gap energy with decreasing NC size. For the hole $g$-factor only small deviations from the universal dependence are expected. The theory results are confirmed by experiments on CsPbI$_3$ NCs. The combination of halide exchange, also considering  mixed compounds like CsPb(I,Br)$_3$, CsPb(Br,Cl)$_3$  or CsPb(I,Cl)$_3$, and NC size and shape variation offers a tool of great flexibility for tailoring the electron and hole $g$-factors and for selecting the desired values for them. We are convinced that very similar trends will be found for the $g$-factors in hybrid organic-inorganic NCs, as the band gap is mostly controlled by the halogen exchange and is only weakly dependent on the cation type.  This is of great importance for both basic research on the spin physics and for spintronic applications of lead halide perovskite NCs.

\section{Methods}\label{sec:meth}

\paragraph{Tight-binding calculations.}
\label{seq:ETB}

To model the quantum confined NC states we use the $sp^3d^5s^*$ tight-binding method. It has been demonstrated that this method permits the precise description of the band structure in inorganic lead halide perovskites~\cite{Nestoklon21}. Moreover, it allows for the realistic description of the surface of these materials: ETB gives a qualitatively correct description of the PbI [001] surface of CsPbI$_3$ without passivation~\cite{Nestoklon21}. In Ref.~\citenum{Kirstein22} this approach was used to demonstrate that the $g$-factors of charge carriers follow simple trends and depend almost exclusively on the band gap energy.

For the empirical tight-binding calculations of the $g$-factors in bulk materials we follow the procedure explained in detail in the Supplementary Information of Ref.~\citenum{Kirstein22}. However, as shown in Ref.~\citenum{Kirstein22}, the values of the bulk $g$-factors calculated within ETB strongly deviate from the experimentally measured $g$-factors. We attribute the origin of this deviation to peculiarities of the DFT calculations underlying the ETB parametrization and to the ETB model itself. First, the modified Becke-Johnson (mBJ) exchange-correlation potential\cite{Tran09,Jishi14} is good to obtain a band gap energy close to experimental data, but it underestimates the renormalization of the energies of the conduction band secondary minima and, thus, the carrier effective masses.\cite{Waroquiers13} Second, for the ETB model used, an accurate description of the halide $p$-bands is hardly possible without inclusion of the interaction between next-nearest neighbors~\cite{Kashikar18}. The lack of such interaction in our model is compensated by the interaction of the halide $p$-states with other bands~\cite{Nestoklon21}, but, as a result, this interaction is overestimated. The effective masses of the charge carriers are correct as long as the dispersion is reasonably well described in the full Brillouin zone, but more subtle properties, including the $g$-factor values, are affected by the overestimated interband interactions. The importance of the ETB parametrization for the $g$-factor values has been discussed in Ref.~\citenum{Jancu05}.

To reach better agreement with the experimental data, we choose a new set of tight-binding parameters to reproduce more precisely the experimentally measured $g$-factors in the bulk crystals CsPb$X_3$, $X=$ I, Br, Cl, at the cost of a less accurate description of the lower halide $p$-bands. The comparison of the band structure calculated using DFT\cite{Tran09,Jishi14,Nestoklon21} and ETB with the new parameters is presented in Supporting Information S1.

\paragraph{\texorpdfstring{${\bf k \cdot p}$}{kp} calculations.}\label{sec:kp}
For qualitative analysis of the ETB calculations we use the \KP\ 8-band approach developed in Ref.~\citenum{Kirstein22}. Note that this approach is valid only for large NCs, where the wave vector associated with the quantum confinement which is inversely proportional to the NC size is small enough to ignore the non-parabolicity of the bands. Details of the \KP\ calculations are presented in Supporting Information S2.

\paragraph{Synthesis of CsPbI\texorpdfstring{$_3$}{3} NCs.} 

The studied CsPbI$_3$ nanocrystals embedded in fluorophosphate Ba(PO$_3$)$_2$-AlF$_3$ glass were synthesized by rapid cooling of a glass melt enriched with the components needed for the perovskite crystallization. The details of the method are given in Refs.~\citenum{Kolobkova2021,Kirstein22_ml}. Samples of fluorophosphate (FP) glass with the composition BaI$_2$-doped 35P$_2$O$_5$--35BaO--5AlF$_3$--10Ga$_2$O$_3$--10PbF$_2$--5Cs$_2$O (mol. \%) were synthesized using the melt-quench technique. The glass synthesis was performed in a closed glassy carbon crucible at the temperature of $T=1050^\circ$C. About 50\,g of the batch was melted in the crucible for 30\,min., then the glass melt was cast on a glassy carbon plate and pressed to form a plate with a thickness of about 2\,mm. Samples with a diameter of 5\,cm were annealed at the temperature of $50^\circ$C below $T_g=400^\circ$C to remove residual stresses. The CsPbI$_3$ perovskite NCs were formed from the glass melt during the quenching. The glasses obtained in this way are doped with CsPbI$_3$ NCs. The dimensions of the NCs in the initial glass were regulated by the concentration of iodide and the rate of cooling of the melt without heat treatment above $T_g$. Three samples were investigated in this paper, which we label \#1, \#2 and \#3. Their technology codes are EK7, EK31 and EK8, respectively. They differ in the NC sizes, which is reflected by relative spectral shifts of their optical spectra. 

The change of the NC size was achieved by changing the concentration of iodine in the melt. Due to the high volatility of iodine compounds and the low viscosity of the glass-forming fluorophosphate melt at elevated temperatures, an increase in the synthesis time leads to a gradual decrease in the concentration of iodine in the equilibrium melt. Thus, it is possible to completely preserve the original composition and change only the concentration of iodine due to a smooth change in the synthesis duration. Glasses with different NC sizes were synthesized using different time intervals. Glasses with the photoluminescence lines centered at  1.801, 1.808 and 1.809~eV (room temperature measurements) were synthesized within 40, 35 and 30 min, respectively. 


\paragraph{Time-resolved Faraday ellipticity (TRFE):} 

The coherent spin dynamics were measured by a pump-probe time-resolved technique~\cite{Yakovlev_Ch6,YugovaPRB09}. We use a titanium-sapphire (Ti:Sa) laser emitting 1.5\,ps long pulses with a spectral width of about $1$\,nm (1.5\,meV) at a pulse repetition rate of 76\,MHz (repetition period $T_\text{R}=13.2$\,ns). The laser photon energy was tunable in the spectral range of $1.265-1.771$\,eV ($700 - 980$\,nm). The laser beam was split into two beams, pump and probe. The probe pulses were delayed with respect to the pump pulses by a mechanical delay line. Both pump and probe beams were modulated using photo-elastic modulators. The pump beam helicity was modulated between $\sigma^+$ and $\sigma^-$ circular polarization at the frequency of 50\,kHz.  The probe beam was kept linearly polarized, but its amplitude was modulated at the frequency of 84\,kHz. The polarization of the transmitted probe beam was detected with a balanced photodiode and analyzed, via a lock-in technique, with respect to the variation of its ellipticity (Faraday ellipticity). 

The measurements were performed at the low temperature of $T=5$\,K, with the sample placed in cooling helium gas. The cryostat was equipped with a vector magnet with three pairs of orthogonally oriented coils, allowing us to apply magnetic fields up to 3\,T in any direction. We used only the Voigt geometry where the magnetic field is perpendicular to the light $k$-vector ($\mathbf{B}_\text{V} \perp \mathbf{k}$). In the transverse magnetic field, the Faraday ellipticity amplitude oscillates in time, reflecting the Larmor spin precession of the carriers. It decays with increasing pump-probe time delay due to spin dephasing. When both electrons and holes contribute to the Faraday ellipticity signal, which is the case for the studied perovskite NCs, the signal can be described as a superposition of two decaying oscillatory functions: $A_{\rm FE} = S_{\rm e} \cos (\omega_{\rm L, e} t) \exp(-t/T^*_{\rm 2,e}) + S_{\rm h} \cos (\omega_{\rm L, h} t) \exp(-t/T^*_{\rm 2,h})$. Here $S_{\rm e(h)}$ are the signal amplitudes that are proportional to the spin polarization of electrons (holes), and $T^*_{\rm 2,e(h)}$ are the carrier spin dephasing times. The $g$-factors are evaluated from the Larmor precession frequency $\omega_{\rm L, e(h)}$ by $|g_{\rm e(h)}|=  \hbar \omega_{\rm L, e(h)}/ (\mu_{\rm B} B)$. Note, that the electron and hole Zeeman splitting is $E_{\rm Z, e(h)}  = \hbar \omega_{\rm L, e(h)}$.

\section*{Acknowledgments}
M.O.N. acknowledges support by TU Dortmund University core funds. E.K. and D.R.Y. acknowledge financial support by the Deutsche Forschungsgemeinschaft via the SPP2196 Priority Program (Project YA 65/26-1). E.A.Zh. and M.B. are thankful to the International Collaborative Research Center TRR160 (Project A1). M.M.G., M.A.S. and E.L.I. acknowledge support of the Russian Foundation for Basic Research (Grant No. 19-52-12038). E.V.K and M.S.K. acknowledge the Saint-Petersburg State University (Grant No. 94030557). 
We thank Volker Brandt from TU Dortmund for the STEM imaging. 
\putbib[perovskite_QDs]
\end{bibunit}

\clearpage
\newpage
\onecolumngrid

\begin{center}
  \textsf{\textbf{\Large Supplementary Information }} \\
\end{center}
\setcounter{equation}{0}
\setcounter{figure}{0}
\setcounter{table}{0}
\setcounter{page}{1}
\setcounter{section}{0}
\makeatletter
\renewcommand{\thepage}{S\arabic{page}}
\renewcommand{\theequation}{S\arabic{equation}}
\renewcommand{\thefigure}{S\arabic{figure}}
\renewcommand{\thetable}{S\arabic{table}}
\renewcommand{\thesection}{S\arabic{section}}
\renewcommand{\bibnumfmt}[1]{[S#1]}
\renewcommand{\citenumfont}[1]{S#1}


\renewcommand{\S}{\mathop{\mathcal S}}
\newcommand{\X}{\mathop{\mathcal X}}
\newcommand{\Y}{\mathop{\mathcal Y}}
\newcommand{\Z}{\mathop{\mathcal Z}}
\begin{bibunit}

\section{Tight-binding parameters}\label{sec:TB_tables}

\begin{figure*}
  \vspace{0.5cm}
  \includegraphics[width=0.8\textwidth]{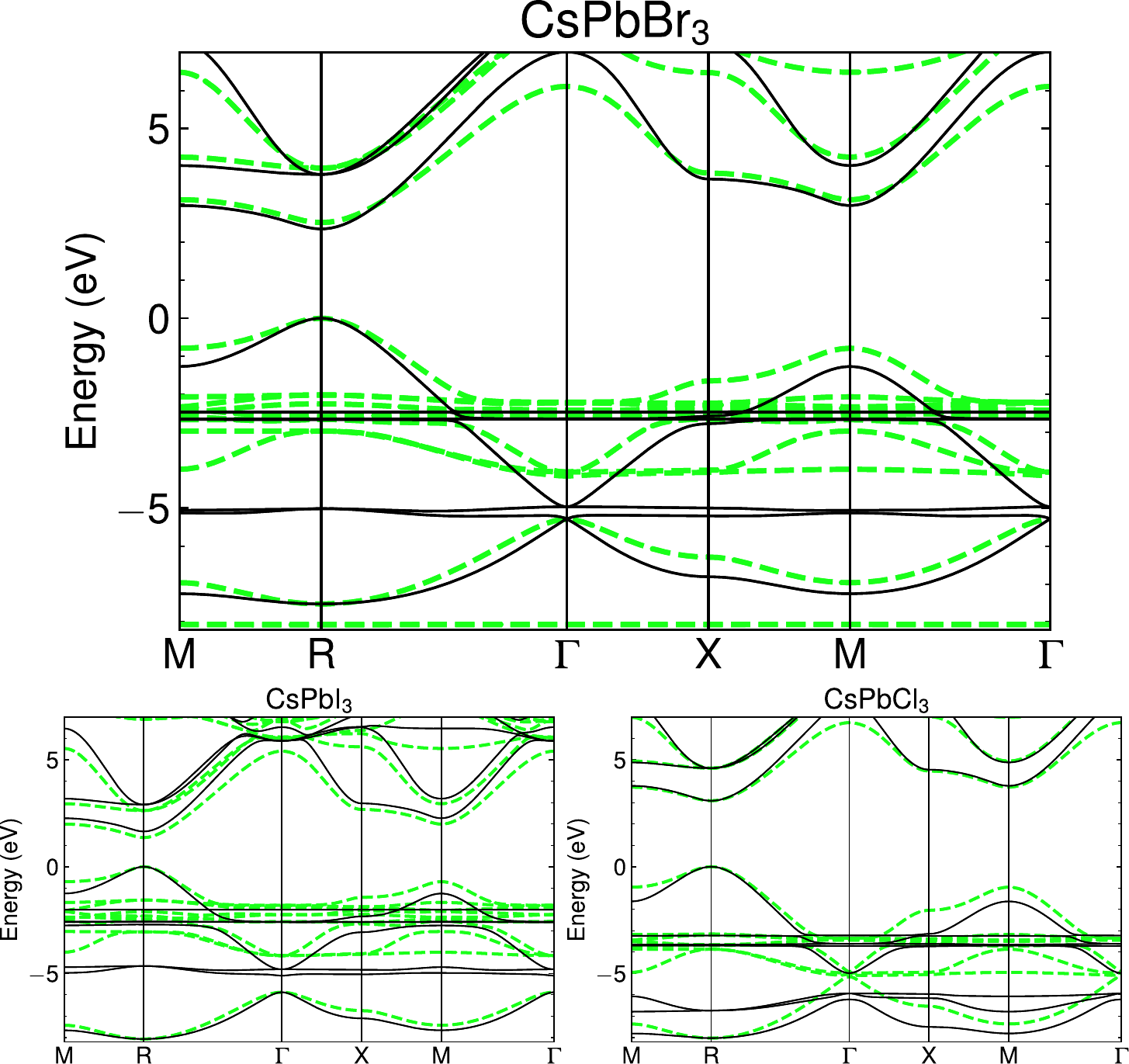}
  \caption{
    Band structure calculated for bulk CsPbI$_3$, CsPbBr$_3$, and CsPbCl$_3$ using ETB with the parameters from Ref.~\citenum{Kirstein22}, corrected in accordance with Table~\ref{tbl:ETB_par_cmp} (black lines) compared with DFT calculations (green dashed lines). For details of the DFT calculations see the Supporting Information of Ref.~\citenum{Kirstein22}.
  }\label{fig:BS} 
\end{figure*}

The ETB parametrization in Ref.~\citenum{Kirstein22} gives an almost perfect description 
of the band structure of bulk cubic CsPb$X_3$, calculated within the DFT approach. However, the resulting bulk $g$-factors of electrons and holes significantly deviate from the experimental data obtained in Ref.~\citenum{Kirstein22}. We attribute this to an incorrect parametrization of the interaction with the halide $p$-band. To improve the modeling of the $g$-factors, we changed the parameters of the interaction between the Pb and the halide atoms, the modified parameters are given in Table~\ref{tbl:ETB_par_cmp}. This change results in a slightly larger difference between the ETB and DFT band structures for CsPb$X_3$ in the cubic phase, but the $g$-factors are much closer to the experimental data. We also changed the diagonal energy of the Pb $p$-orbital for CsPbI$_3$ and CsPbBr$_3$ to match the low-temperature experimental band gap of MAPbI$_3$ and CsPbBr$_3$ from Ref.~\citenum{Kirstein22}. The band structure, compared with DFT calculations, is shown in Fig.~\ref{fig:BS}. The values of the carrier effective masses and $g$-factors are given in Table~\ref{tbl:kp_from_ETB}.


\begin{table}[bh]\caption{Modified ETB parameters used in the calculations. We give only the parameters which differ from the parameters presented in Ref.~\citenum{Kirstein22}. All values are given in eV.}
\label{tbl:ETB_par_cmp}
 \begin{tabular*}{\linewidth}{@{\extracolsep{\fill}}lrrr}
 \hline
 \hline
 &   CsPbI$_3$  &  CsPbBr$_3$   & CsPbCl$_3$  \\
\hline
$         E_{pc}$  &  $    4.81$& $    5.39$  &    $ 6.46$    \\
$       pp\sigma$  &  $   -2.47$& $   -2.61$  &    $-2.79$    \\
$          pp\pi$  &  $    0.00$& $    0.05$  &    $ 0.20$    \\
$   p_ad_c\sigma$  &  $    2.62$& $    2.90$  &    $ 3.51$    \\
 \hline
 \hline
 \end{tabular*}
 \end{table} 



\begin{table}[bh]\caption{Effective masses and $g$-factors of electrons and holes calculated for the bulk materials using the parameters from Ref.~\citenum{Kirstein22}, corrected in accordance with Table~\ref{tbl:ETB_par_cmp}.}
\label{tbl:kp_from_ETB}
\begin{tabular*}{\linewidth}{@{\extracolsep{\fill}}lrrr}
 \hline
 \hline
                &   CsPbI$_3$  & CsPbBr$_3$    & CsPbCl$_3$ \\
\hline
$E_g$ (eV)      &  $ 1.652$  &  $ 2.352$   & $3.090$ \\
$\Delta$ (eV)   &  $ 1.258$  &  $ 1.436$   & $1.526$ \\
$m_{\rm e}/m_0$ &  $ 0.168$  &  $ 0.219$   & $0.315$ \\
$m_{\rm h}/m_0$ &  $ 0.145$  &  $ 0.191$   & $0.250$ \\
$g_{\rm e}$     &  $+3.23 $  &  $+1.77 $   & $+0.95 $ \\
$g_{\rm h}$     &  $-0.33 $  &  $+0.66 $   & $+1.14 $ \\
 \hline
 \hline
 \end{tabular*}
 \end{table} 

The band gaps agree well with the combined experimental and theoretical (DFT) studies on bulk crystals giving 1.72\,eV, 2.31\,eV, and 2.99\,eV for CsPb$X_3$ in Ref.~\citenum{Tao2019}.


\section{\KP\ model}\label{sec:app:kp}

To describe the electron and hole $g$-factor renormalization within the \KP~approach, we consider the 8 band model that includes the two-fold degenerate valence band states and the six states in the conduction band: the bottom two-fold degenerate (spin-orbit split) conduction band and the higher band that is four-fold degenerate at the $R$ point of the Brillouin zone and consisting of the heavy-electron and light-electron branches.

The Bloch amplitudes at the $R$ point are taken in the form\\
\begin{subequations}
\label{Bloch:amp}
valence band:
\begin{equation}
\label{Bloch:v}
\begin{cases}
u_{vb,1/2}(\bm r) = \mathrm i \mathcal S(\bm r)|\uparrow\rangle,\\
u_{vb,-1/2}(\bm r) =\mathrm i \mathcal  S(\bm r)|\downarrow\rangle,\\
\end{cases}
\end{equation}
bottom conduction band:
\begin{equation}
\label{Bloch:c}
\begin{cases}
u_{cb,1/2}(\bm r) = -\sin{\vartheta} \Z(\bm r)|\uparrow\rangle - \cos{\vartheta} \cfrac{\X(\bm r) + \mathrm i \Y(\bm r)}{\sqrt{2}}|\downarrow\rangle,\\
u_{cb,-1/2}(\bm r) =+\sin{\vartheta} \Z(\bm r)|\downarrow\rangle - \cos{\vartheta} \cfrac{\X(\bm r) - \mathrm i \Y(\bm r)}{\sqrt{2}}|\uparrow\rangle,\\
\end{cases}
\end{equation}
split-off c.b. (light electron):
\begin{equation}
\label{Bloch:cl}
\begin{cases}
u_{le,1/2}(\bm r) = \cos{\vartheta} \Z(\bm r)|\uparrow\rangle - \sin{\vartheta} \cfrac{\X(\bm r) + \mathrm i \Y(\bm r)}{\sqrt{2}}|\downarrow\rangle,\\
u_{le,-1/2}(\bm r) =\cos{\vartheta} \Z(\bm r)|\downarrow\rangle + \sin{\vartheta} \cfrac{\X(\bm r) - \mathrm i \Y(\bm r)}{\sqrt{2}}|\uparrow\rangle,\\
\end{cases}
\end{equation}
split-off c.b. (heavy electron):
\begin{equation}
\label{Bloch:ch}
\begin{cases}
u_{he,3/2}(\bm r) = -\cfrac{\X(\bm r) + \mathrm i \Y(\bm r)}{\sqrt{2}}|\uparrow\rangle,\\
u_{he,-3/2}(\bm r) = \cfrac{\X(\bm r) - \mathrm i \Y(\bm r)}{\sqrt{2}}|\downarrow\rangle.\\
\end{cases}
\end{equation}
\end{subequations}
Here $\uparrow$ and $\downarrow$ denote the basic spin-$1/2$ spinors, the function ${\cal S}(\bm r)$ is invariant in the group Pm$\bar{3}$m, and the functions ${\cal X}(\bm r), {\cal Y}(\bm r), {\cal Z}(\bm r)$ transform like the corresponding coordinates, the subscripts $vb$, $cb$, $le$, and $he$ refer, respectively, to the \emph{valence band}, \emph{(bottom) conduction band}, spin-orbit split-off \emph{light electron}, and spin-orbit split-off \emph{heavy electron}. For the two latter bands the notations $h$ and $l$ are used in the same way as for the holes in III-V or II-VI semiconductors.
In what follows we focus on the perovskite cubic phase where $\cos{\vartheta} = \sqrt{2/3}$, $\sin{\vartheta} = \sqrt{1/3}$, and the space group of symmetry is $Pm\bar{3}m$ (or O$_h^1$). 
At the $R$ point, the light and heavy split-off electrons are degenerate forming a quadruplet with total angular momentum $3/2$.
The representations at the $R$ point are $R_6^-$ and $R_8^-$ (corresponding to the representations $\Gamma_6^-, \Gamma_8^-$ of the point group $O_h$ in the Koster notation\cite{Koster_book}) for the conduction bands and $R_6^+$ (or $\Gamma_6^+$) for the valence band. We define the band gap $E_g$ as the energy difference between the bottom conduction band and the top valence band, and let $\Delta>0$ be the spin-orbit splitting between the two-fold and the four-fold conduction bands. The interband momentum matrix element is defined by \begin{equation}
\label{pcv}
p = \mathrm i\langle \Z |\hat{p}_z|\S\rangle =  \mathrm i \langle \X |\hat{p}_x|\S\rangle =  \mathrm i \langle \Y |\hat{p}_y|\S\rangle\,.
\end{equation}
Here $\hat{p}_{x,y,z}$ are the components of the momentum operator, and we take the phases of the Bloch functions in such a way that $p$ is real. 

We start the analysis of the $g$-factors with the case of the holes. Following Ref.~\citenum{Kirstein22} we obtain for a bulk crystal of cubic symmetry
\begin{equation}
\label{gh}
g_{\rm h} = 2- \frac{4}{3}\frac{p^2}{m_0}\left(\frac{1}{E_g} - \frac{1}{E_g+\Delta}\right),
\end{equation}
where $m_0$ is the free electron mass. The second contribution is related to the magnetic-field-induced \KP-mixing with the bottom conduction band. The third term is related to the mixing with the higher 4-fold degenerate conduction band. The remote band contributions in the case of holes are rather small.\cite{Kirstein22} Within the \KP-approach for simplicity of the analysis we consider the size-quantization effect in a NC of spherical symmetry and expect a relatively weak effect of the NC shape on the $g$-factors, cf. Ref.~\citenum{Avdeev23}. Kiselev et al.~\citenum{Kiselev98} derived in the \KP~method the following equation for the electron $g$ factor in a spherical quantum dot, composed of III-V based semiconductor material A embedded in the wider band gap matrix of material B: 
\begin{eqnarray} 
g &=& 2 +  [g_{\rm A}(E_{\mathrm e}) -2]\: w_{\rm A} + [g_{\rm
B}(E_{\mathrm e}) -2]\: w_{\rm B}  \\ &&+\:[g_{\rm B}(E_{\mathrm e}) - g_{\rm
A}(E_{\mathrm e})]\:V_{QD}(R)\:f^2(R) \:. \nonumber
\end{eqnarray}
Here $R$ and $V_{QD}(R) = 4 \pi R^3/3$ are the radius and volume of the quantum dot, $f(r)$ is the conduction-electron scalar envelope, $g_{\rm A}(E)$ and $g_{\rm B}(E)$ are defined by Eq. (3) of the main text, where the energy $E_{\mathrm h}$ is replaced by the electron confinement energy $E_{\mathrm e}$, $w_{\rm A}$ and $w_{\rm B}$ are the integrals $\int d{\bm r}f^2(r)$, taken over the A and B volumes, respectively. Importantly, the sum $w_{\rm A}+ w_{\rm B}$ differs from unity because of the confinement-induced conduction band-valence band mixing~\cite{Kiselev98,ivchenko05a}. If the potential barriers are high, the values of $f^2(R) $ and $w_{\rm B}$ become negligible, and the $g$ factor tends toward $2+w_{\rm A} [g_{\rm A}(E_{\rm e})-2]$. Since the role of the conduction band in III-V based semiconductors is in effect taken by the valence band in perovskites, the hole $g$-factor in the perovskite NC is given by 
\begin{equation}
\label{gh:NC}
g_{\rm h}(E_{\rm h}) = 2- \frac{4}{3}\frac{p^2}{m_0}w_{\rm h}\left(\frac{1}{E_{\rm h}+ E_g} - \frac{1}{E_{\rm h}+ E_g+\Delta}\right).
\end{equation}
Here $E_h$ is the hole quantization energy, and the factor $w_h$ is given by  
\begin{equation}\label{eq:ws}
w_{\rm h}(E) = \frac{1}{1+ \frac{2m(E)c^2(E)}{\hbar^2}E}\:,
\end{equation}
where
\[
\frac{1}{m(E)} = \frac{2}{3} \frac{p^2}{m_0^2} \left(\frac{2}{E_g+E} + \frac{1}{E_g+\Delta+ E} \right)
\]
and
\[
c^2 (E) = \frac{ \hbar^2 p^2}{3m_0^2}\left(\frac{2}{(E_g + E)^2} + \frac{1}{(E_g+\Delta + E)^2}\right)\:.
\] 
This is in agreement with Eq. (3) of the main text. 

\begin{figure*}
\includegraphics[width=0.8\textwidth]{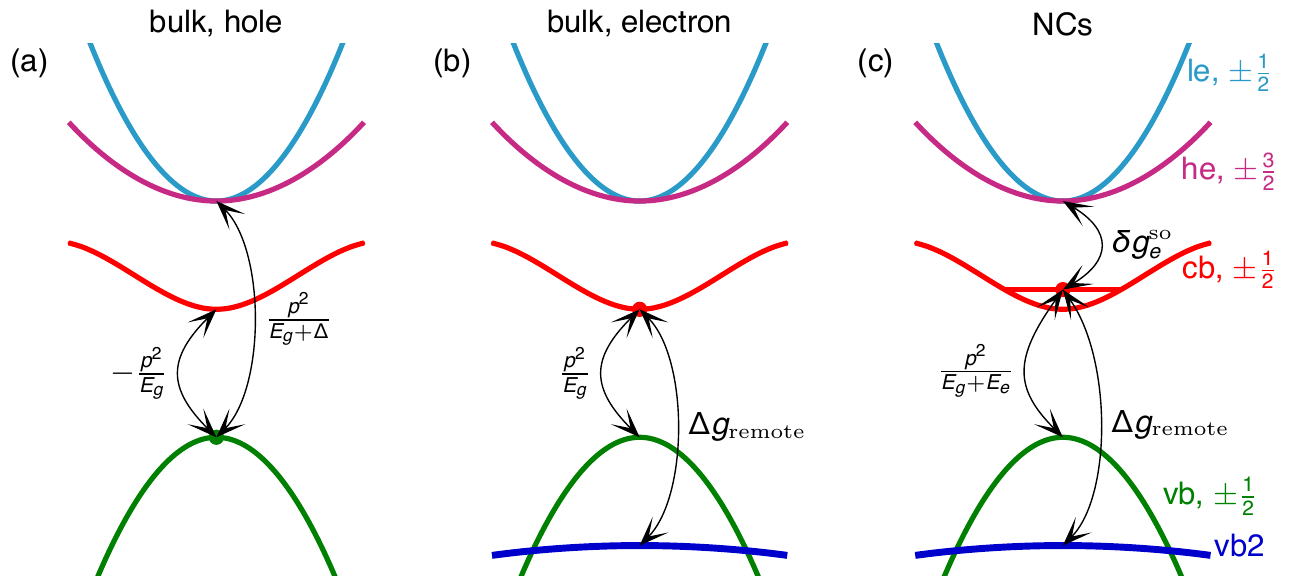}
\caption{Illustration of the bands important for the calculation of: (a) hole $g$-factor in bulk perovskites,\cite{Kirstein22} Eq.~\eqref{gh} (b) electron $g$-factor in bulk perovskites,\cite{Kirstein22} Eq.~\eqref{ge_bulk}, and (c) electron $g$-factor in perovskite NCs, Eq.~\eqref{ge}. 
}
\label{fig:g_illustr}
\end{figure*}

Now we turn to the case of the conduction band electrons. 
The analysis, following the same lines as above, but starting from the bulk electron $g$-factor\cite{Kirstein22},
\begin{equation}
\label{ge_bulk}
g_{\rm e} = -\frac{2}{3} + \frac{4}{3} \frac{p^2}{m_0}\frac{1}{E_g} + \Delta g_{\rm remote}\,,
\end{equation}
yields the expression for the NCs:
\begin{equation}
\label{ge}
g_{\rm e}(E_{\rm e}) = -\frac{2}{3} + \frac{4}{3} \frac{p^2}{m_0}\frac{{w_{\rm e}}}{E_g+E_{\rm e}} + \Delta g_{\rm remote} + \delta g_{\rm e}^{\rm so}\,.
\end{equation}
Here the term $-2/3$ comes from the spin structure of the Bloch amplitudes,\footnote{For bulk crystals of tetragonal symmetry the electron and hole $g$-factors are anisotropic\cite{Kirstein22}. The principal components of the heavy electron $g$-factors tensor $ g_\parallel \equiv g_{zz}$ and $g_\perp\equiv g_{xx} = g_{yy}$ with $z$ being the $C_4$-axis are
\[
g_{e\parallel} = 2(\sin^2{\vartheta}-\cos^2{\vartheta}) + \frac{2p_\perp^2}{m_0} \frac{\cos^2{\vartheta}}{E_g}, \quad g_{e\perp} =-2\sin^2{\vartheta} + \frac{2\sqrt{2}p_\parallel p_\perp}{m_0} \frac{\cos{\vartheta}\sin{\vartheta}}{E_g},
\]
where $p_\parallel \equiv \mathrm i\langle \Z |\hat{p}_z|\S\rangle$, $p_\perp \equiv  \mathrm i \langle \X |\hat{p}_x|\S\rangle =  \mathrm i \langle \Y |\hat{p}_y|\S\rangle$. These expressions correct typos in Eqs.~(S12) of Ref.~[\citenum{Kirstein22}].} 
the second term results from the conduction-valence band \KP-mixing with $E_e$ being the electron quantization energy, $w_e$ is an energy-dependent coefficient, and $\delta g_{\rm e}^{\rm so}$ is contributed by the higher conduction band $R_8^-$. The band contributions to the resulting $g$-factors in bulk materials and NCs are illustrated in Fig.~\ref{fig:g_illustr}. In order to get a reasonable estimate, we first consider the two-band approximation excluding the band $R_8^-$ as if $\Delta \gg E_g$. In this model the energy spectrum in absence of the magnetic field is symmetric, $w_e = w_h$, and the electron and hole quantization energies coincide, $E_e = E_h$. Leaving only the linear in $E_{e/h}$ terms in \eqref{eq:ws} in the limit $\Delta\to\infty$, we obtain
\begin{subequations}
\begin{equation}
\label{gh:NC:1}
g_{\rm h}(E_{\rm h}) \approx 2- \frac{4}{3}\frac{p^2}{m_0}\frac{1}{2E_{\rm h}+ E_g}.
\end{equation}
\begin{equation}
\label{ge:NC:1}
g_{\rm e}(E_{\rm e}) = -\frac{2}{3} + \frac{4}{3} \frac{p^2}{m_0}\frac{1}{2E_{\rm e}+E_g} + \Delta g_{\rm remote} + \delta g_{\rm e}^{\rm so}\,.
\end{equation}
\end{subequations}

Finally, we include the split-off conduction band in the considerations. Importantly, its presence results in a strong renormalization \cite{ivchenko05a} of the electron $g$-factor described by the last term in Eq.~\eqref{ge_bulk}. We derive an analytical formula for $\delta g_{\rm e}^{\rm so}$ for a spherically-symmetric NC assuming that the electron quantization energy $E_{\rm e}\ll \Delta$. For simplicity, we also assume the inequality $\Delta \ll E_g$. 
Table \ref{78} presents the 2$\times$4 matrix ${\cal H}_{6c,8c}({\bm k})$ of the off-diagonal matrix elements that mix the conduction bands $R_6^-$ and $R_8^-$ in the extended \KP\ approach. The matrix is similar to that describing the coupling between the valence bands $\Gamma_7$ and $\Gamma_8$ in III-V semiconductors \cite{BirPikus,PikusTitkov,Winkler}. Standard notations are used, namely,
\begin{eqnarray}  \label{HIGF}
&& H = \frac{\sqrt{3} \hbar^2 \gamma_3}{m_0} k_z(k_x - {\rm i} k_y)\:,\\ && I = \frac{\sqrt{3} \hbar^2}{2 m_0} \left[ \gamma_2 \left(k_x^2 - k_y^2 \right) - 2{\rm i} \gamma_3 k_x k_y \right]  \:, \nonumber \\&& G - F =  \frac{\hbar^2 \gamma_2}{m_0} \left( k_x^2 + k_y^2 - 2 k_z^2 \right)\:,\nonumber
\end{eqnarray}
and $\gamma_2, \gamma_3$ are the dimensionless Luttinger band parameters, applied here for the complicated perovskite conduction band \cite{Luttinger55}. In the following we ignore the difference between $\gamma_2$ and $\gamma_3$, and use the notation $\bar{\gamma}$ for them. In this spherical approximation $$I = \frac{\sqrt{3} \hbar^2 \bar{\gamma}}{2 m_0} (k_x - {\rm i} k_y)^2\:.$$ 

\begin{center}
\begin{table}\caption{The off-diagonal components of the conduction-electron Hamiltonian ${\cal H}_{6c,8c}$.}
\label{78}
\begin{tabular}{l|l|l|l|l|}
& $he, +3/2$ & $le, +1/2$ & $le, -1/2$ & $he, -3/2$ \\ \hline 
$cb, +1/2$ & $H^*/\sqrt{2}$ & $(G-F)/\sqrt{2}$ & $- \sqrt{3/2}H$ & $- \sqrt{2}I$ \\ \hline 
$cb, -1/2$  & $\sqrt{2}I^*$ & $- \sqrt{3/2}H^*$ & $-(G-F)/\sqrt{2}$ & $H/\sqrt{2}$ \\ \hline
\end{tabular}
 \end{table}
\end{center}

In the presence of a magnetic field ${\bm B} = {\bm \nabla} \times {\bm A}$, the electron wave vector in Eqs.~(\ref{HIGF}) should be replaced by 
\begin{equation}
{\bm K} = {\bm k} - \frac{e}{\hbar c} {\bm A} = - {\rm i} {\bm \nabla} - \frac{e}{\hbar c} {\bm A}\:
\end{equation}
and $(k_x - {\rm i} k_y)$ by $(K_x - {\rm i} K_y)$. We evaluate the spin splitting of the conduction band electron in a NC for $\bm B \parallel z$, making use of second-order perturbation theory as follows
\begin{equation}
\begin{split}
\delta g_{\rm e}^{so} \mu_B B_z = - \frac{1}{\Delta} \big[ & \langle cb, 1s, +1/2 | {\cal H}_{6c,8c}
{\cal H}_{8c,6c}  | cb, 1s, +1/2\rangle  
\\ & -  \langle cb, 1s, -1/2 | {\cal H}_{6c,8c}
{\cal H}_{8c,6c}  | cb, 1s, -1/2\rangle \big] \:.
\end{split}
\end{equation}
Here, $ | cb, 1s, \pm 1/2 \rangle$ is the Kramers conjugate pair of the zero-dimensional electron ground state $1s$, ${\cal H}_{8c,6c} = {\cal H}_{6c,8c}^{\dag}$, and we assume the spin-orbit splitting $\Delta$ to exceed by far the size-quantization energies of the states contributing to $\delta g_{\rm e}^{so}$.

Extracting the $B_z$-linear contributions by virtue of the relations
  \begin{equation} \begin{split}
& k_xk_y - k_yk_x = \frac{\mathrm i e}{\hbar c} B_z,
\\
& (k_x + \mathrm ik_y)(k_x - \mathrm i k_y) \to \frac{eB_z}{\hbar c}, \\
& (k_x + \mathrm ik_y)^2(k_x - \mathrm i k_y)^2 \to 4\frac{eB_z}{\hbar c}(k_x^2+k_y^2),
\end{split}\end{equation}
we arrive at
\begin{equation}
\label{genc}
\delta g_{\rm e}^{\rm so} = {-60 \frac{\hbar^2}{m_0} \frac{\langle k_z^2\rangle}{\Delta} \bar\gamma^2} \,.
\end{equation}
Here $\langle k_z^2\rangle =\langle k_x^2 \rangle = \langle k_y^2 \rangle$ is the average value of $k_z^2$ over the $1s$ electron envelope function. Note that for a spherically-symmetric or cubic NC $\hbar^2 \langle k_z^2\rangle /2m_{\rm e} = E_k/3$,  where $E_k$ is the contribution of the kinetic energy to the electron confinement energy $E_{\rm e}$. For a NC with infinite barriers $E_k = E_{\rm e}$. Equation~(\ref{genc}) can be recast as
\begin{equation}
\label{genc:1}
\delta g_{\rm e}^{\rm so} =  -40 \frac{\bar\gamma^2}{\gamma_1} \frac{E_k}{\Delta}\:,
\end{equation}
where we took into account that $E_k= 3 \hbar^2\gamma_1 \langle k_z^2\rangle/(2m_0)$.

To demonstrate the significance of the interaction with the split-off band we present rough estimates within the Kane two-band model, where the Luttinger parameter $\bar\gamma = p^2/(3m_0E_g)$ and $m_0/m_{\rm e} = 2\bar\gamma (=\gamma_1)$. From Eq.~\eqref{genc:1} we obtain
\begin{equation}
\label{genc:simple}
\delta g_{\rm e}^{\rm so} \approx -20 \bar{\gamma} \frac{E_{\rm e}}{\Delta}.
\end{equation}
Even for $E_{\rm e} \sim 0.1\Delta$ the split-off band admixture contribution $\delta g_{\rm e}^{\rm so} $ is on the order of unity due to the large prefactor.

\begin{figure*}
\includegraphics[width=0.7\textwidth]{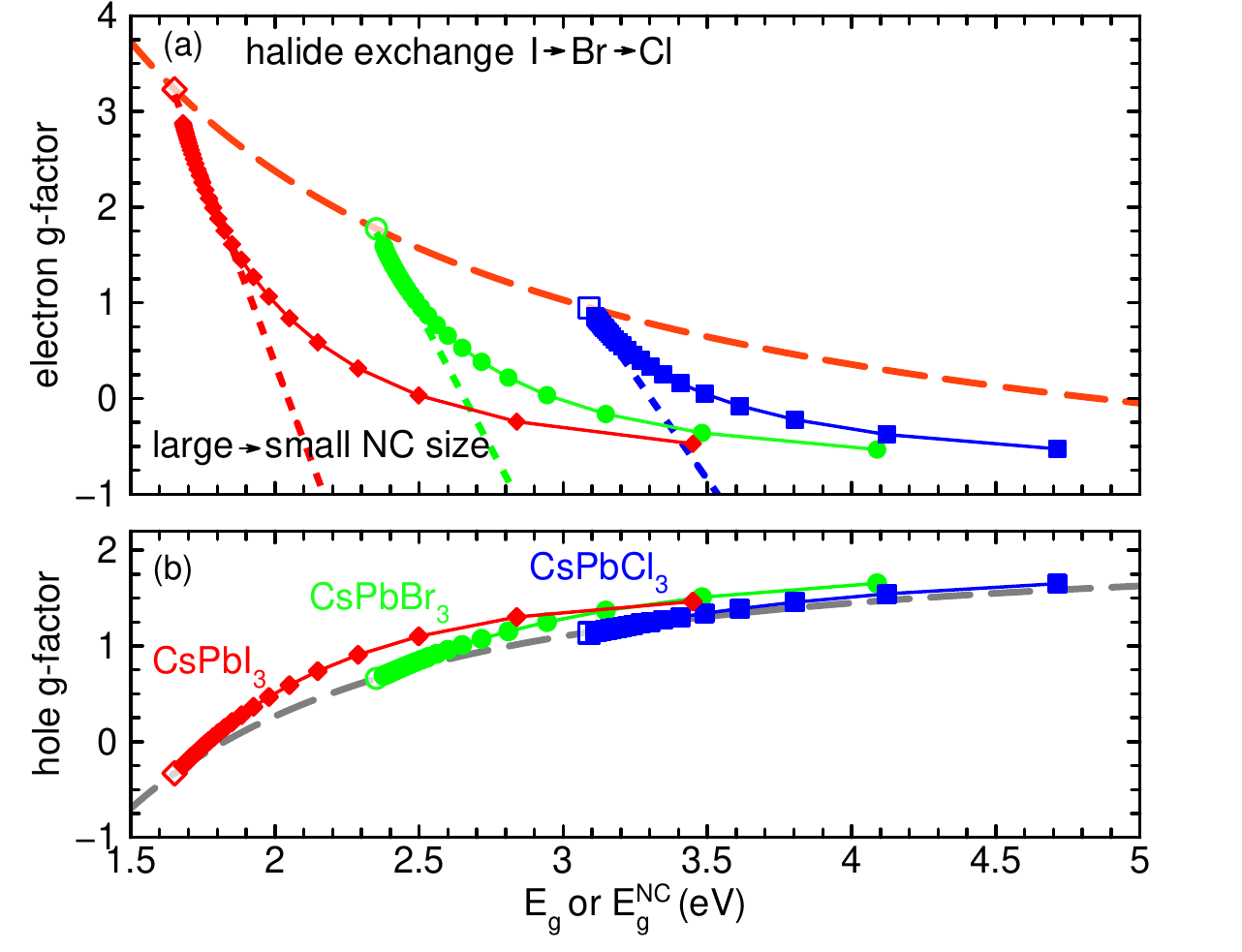}
\caption{Closed symbols show the $g$-factors of electrons (upper panel) and holes (lower panel) calculated for CsPb$X_3$ ($X =$ I, Br, Cl) NCs in ETB. Open symbols show the $g$-factors (from an atomistic approach) in the bulk crystals. Dashed lines show the result of the \KP-approach. For the holes we disregard the effect of size quantization.}
\label{fig:comp}
\end{figure*}

Figure~\ref{fig:comp} shows the comparison of the electron $g$-factors calculated after Eqs.~\eqref{ge}, \eqref{genc:simple} shown by the dashed lines with the ETB results. We used the same value of $p$ (see caption of Figure~\ref{fig:comp}), which reproduces the $g$-factors in the bulk material, and we took into account that the electron quantization energy is the same as the hole quantization energy due to comparable effective masses. One can see that the analytical expressions derived in the \KP-model describe the initial rapid drop of the electron $g$-factors with increasing confinement. Further, the saturation-like behavior is beyond the contribution of first-order in $E_{\rm e}$, Eqs.~\eqref{genc} -- \eqref{genc:simple}.


\section{Temperature dependence of the band gap}

\begin{figure}%
\includegraphics[width=0.5\textwidth,trim={1.4cm 0 2.8cm 0},clip]{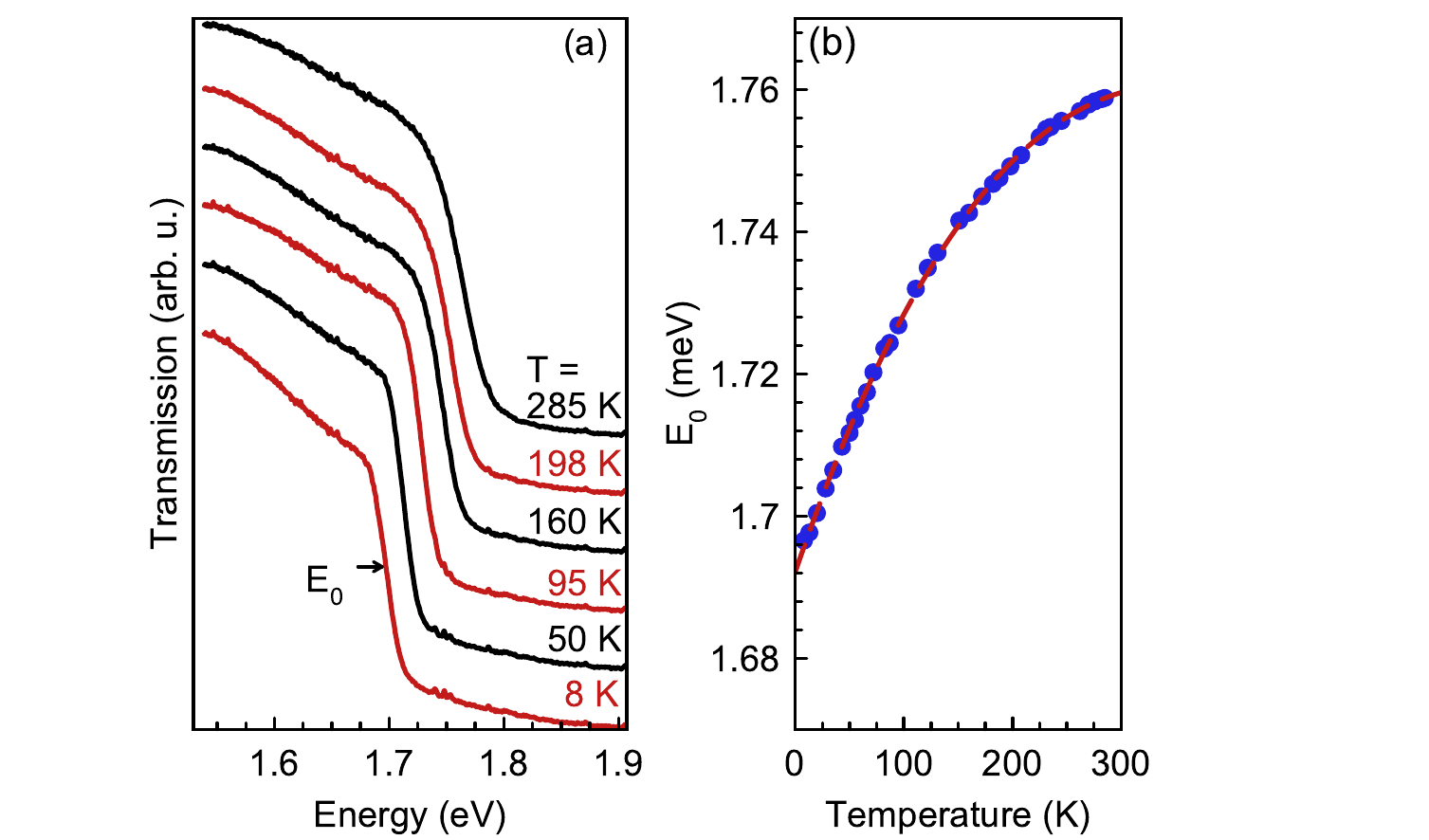}
\caption{(a) Transmission spectra of CsPbI$_3$ NCs (sample \#1) measured from cryogenic to room temperature. The spectra are shifted vertically for clarity. The arrow indicates the center of the curve $E_0$.
(b) Temperature dependence of $E_0$. The line shows the fit with Eq.~\ref{eq:Temperature}. }%
\label{fig:temperature_SI}%
\end{figure}

We measured transmission spectra with a halogen lamp and a 0.5~m monochromator with an attached charge-coupled devices camera. The spectra measured for the sample \#1 in the temperature range from 8 up to 285~K are shown in Fig.~\ref{fig:temperature_SI}(a). The transmission has a step like decrease around 1.7~eV, which reflects the strong light absorption above the band gap. 


With increasing temperature the CsPbI$_3$ NCs' central absorption energy $E_0$ shows a monotonic high energy shift by 60~meV. The shift is linear for temperatures below 100~K and gradually saturates with further temperature increase. The slope of the edge also increases with rising temperature, it is almost linear for temperatures above 50\,K but shows saturation dependence for low temperatures. In perovskites, the lattice shrinkage dependent gap change is far stronger than electron-phonon interaction (Varshni-term), which acts in the opposite energy direction. We assume a linear shrinkage of the lattice constant with temperature, resulting in a linear dependence of the band gap on temperature~\cite{Zhang2020,Yu2021}. Together, it reads
\begin{equation} \label{eq:Temperature}
E_0(T) = E_{0}(T=0)-\underbrace{\frac{\alpha T^2}{T+\beta}}_{\rm Varshni}+\underbrace{\zeta T}_{\rm lattice~shrinkage}
\end{equation}
for describing the band gap shift. We fit the experimental dependence with the parameters $\zeta= 0.447$~meV/K as the linear slope, $E_{g}(T=0)=1.692$~eV  as the zero temperature band gap, and the Varshni parameters $\alpha=1.2$~meV/K and $\beta = 1300$~K. 



\section{Scanning transmission electron microscopy -- high-angle annular dark-field}

\begin{figure*}%
\includegraphics[width=\textwidth]{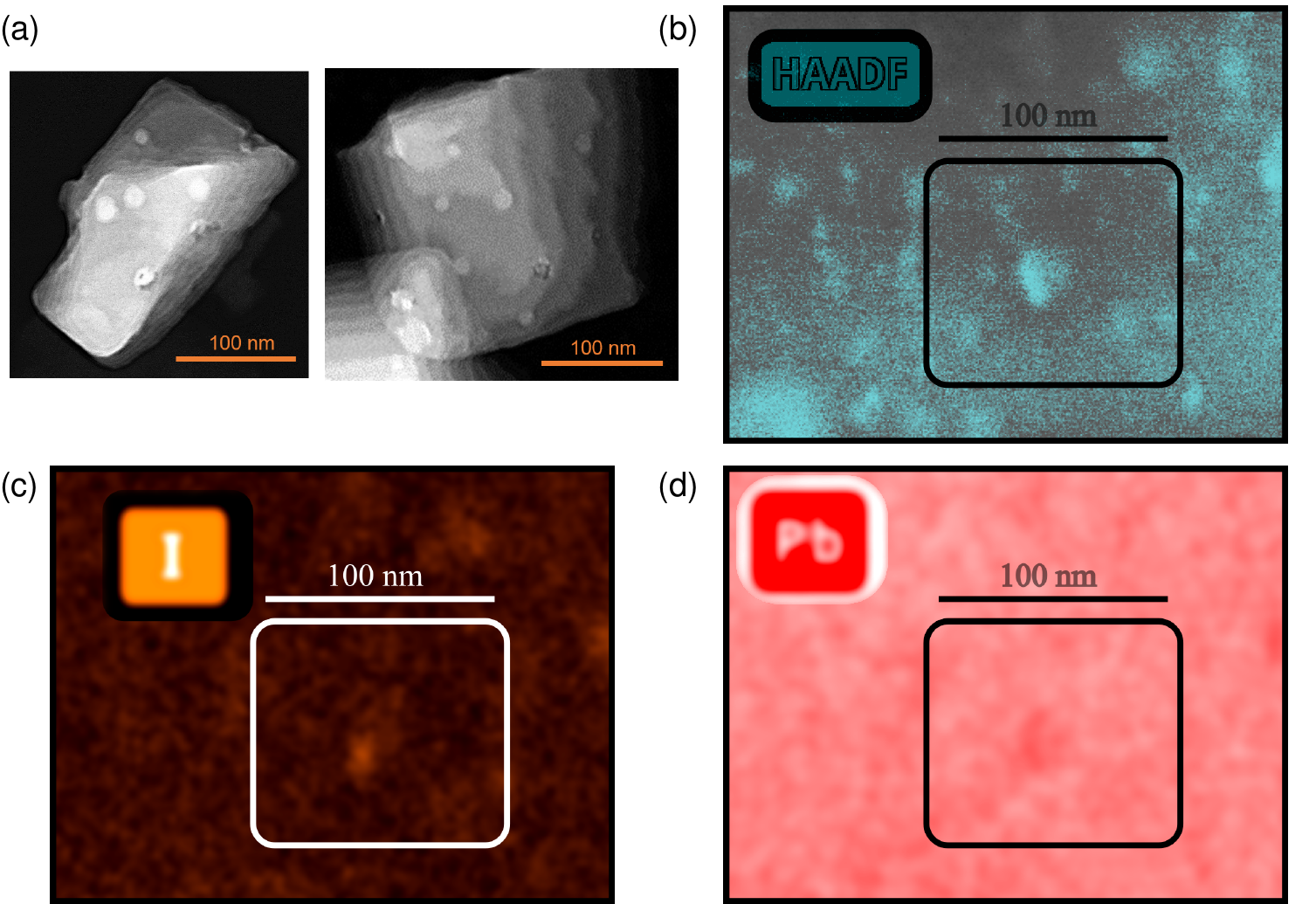}
\caption{(a) STEM-HAADF images of glass grains with CsPbI$_3$ NCs (sample \#1) measured at room temperature. The bright spots are NCs. The plateau-like brightness steps result from the sharpening process to better identify the NCs.  
(b) STEM-HAADF image of another grain which was analyzed for elemental composition in (c-d). 
(c,d) Detected iodine (c) and lead (d) signal within the sample region presented in (b).
}
\label{fig:TEM_elemental}%
\end{figure*}

The evaluation of the NC sizes was done  by means of scanning transmission electron microscopy using the high-angle annular dark-field method (STEM-HAADF). For these measurements the samples were grinded with an agate mortar. The powder was placed on a carbon coated copper grid. The TEM images were recorded using a Talos F200X machine of Thermo Fisher with the acceleration voltage of 200\,kV, current of 130\,pA and high-angle annular dark field detector (0.16\,nm resolution) [1024$\times$1024 pixel Thermo-Fisher SuperX]. Due to the dielectric glass matrix in which the CsPbI$_3$ NCs are embedded, charges can easily accumulate on the samples and, therefore, the scanning fails due to distraction of the electron beam. The best results can be achieved on the edges of the powder grains. Overall, on several grains dozens of sharp or blurry bright spots can be identified as the CsPbI$_3$ NCs, e.g. shown in Fig.~\ref{fig:TEM_elemental}(a). Typically the NCs remain blurry, thus the real morphology is hidden but from the sharp NC images a non-cuboid almost spherical shape can be found. 

An elemental analysis was performed to confirm the presence of CsPbI$_3$ NCs, Fig.~\ref{fig:TEM_elemental}(b-d). Interestingly, the analysis shows a rather homogeneous distribution of all elements over the full grain. Noteworthy, within the presented cut of the image the full area shows the presence of the elements (black/white for the absence). Within the glass melt still a high amount of the perovskite-forming elements are solved. However, for iodine and lead at some spots a higher concentration of the elements can be seen. These spots of high concentration coincide with the positions where before the NCs were identified, thus confirming the presence of NCs. Note that in order to have a better contrast, the images were post-processed with the graphic software inkscape. The brightness to contrast ratio and color were tuned, further a slight blurring was applied and image stacking was used. The rough estimation of the NC density within the 2D images is $2\times 10^{10}$~cm$^{-2}$. A precise recalculation to the 3D density is not possible due to the unknown NC thickness and hidden NCs at higher depth. However, we obtain an estimate of ~$3\times10^{6}$ measured NCs within the $200\,\mu$m laser spot. 

\begin{figure*}%
\includegraphics[width=\textwidth]{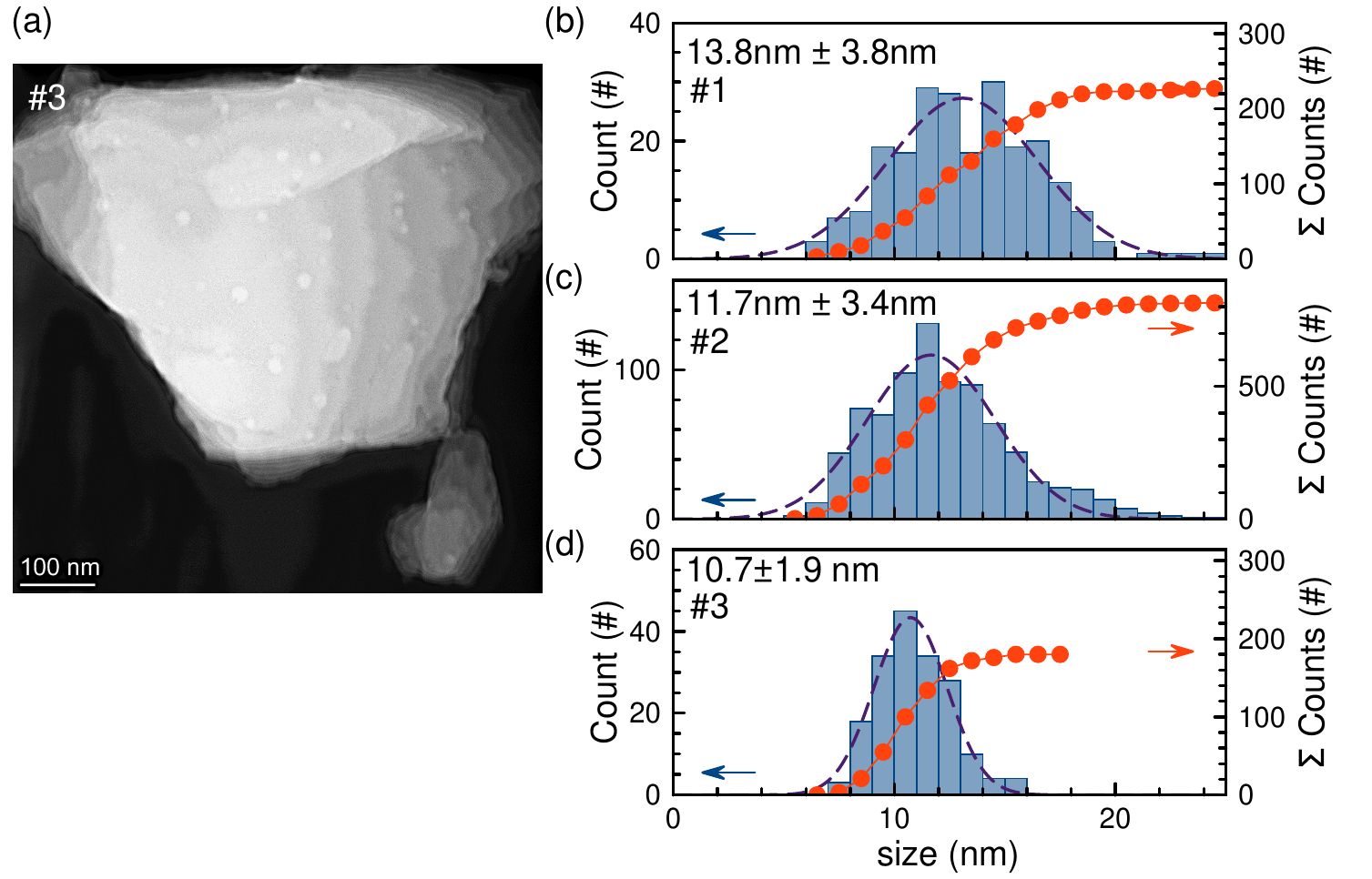}
\caption{(a) STEM-HAADF image of a glass grain with CsPbI$_3$ NCs (sample \#3).
(b-d) Histograms of the NC size distribution (diameter) for samples \#1, \#2 and \#3, respectively (left axis), binned by 1\,nm; the right axis shows the sum of NC counts for specific sizes.
}
\label{fig:TEM_histo}%
\end{figure*}

Finally, for a larger set of measured grains the NC size was evaluated using the calibrated ruler. The size distribution is shown in the histograms in Fig. \ref{fig:TEM_histo}(b-d). Another example of a STEM image for sample \#3 with smaller NCs is shown in Fig.~\ref{fig:TEM_histo}(a). For evaluation of the NC size the software ImageJ, and partially contrast post-processing were used. For the image-post processing the vanilla STEM images are sharpened by a Microsoft PowerPoint algorithm and further the brightness and contrast settings adjusted by ImageJ. Afterwards the NC size was evaluated by an approximation via two lines placed manually inside the NC to take into account the slight elliptical shape of the NCs. In total about 1000 NCs were evaluated. Overall, NCs with sizes varying from 8 up to 16~nm were found. The histograms were fitted assuming Gaussian distributions, neglecting the asymmetry of having more larger than smaller NCs, from which average sizes of 13.8~nm (sample \#1), 11.7~nm (sample \#2), and 10.7~nm (sample \#3) were evaluated. A certain trend of having a broader size distribution for a larger average size is present, originating from difficulties in the growth control.

\section{Lattice constant of cubic materials}\label{sec:latt_const}
In ETB we use the lattice constant of cubic materials. We comment on these values as the corresponding data presented in different publications are not fully consistent. This is mainly explained by the fact that bulk cubic perovskites are not stable at low temperatures. In Ref.~\citenum{Jishi14},  for cubic CsPbI$_3$, the high-temperature experimental value $0.6289$~nm from Ref.~\citenum{Trots08} is used and for cubic CsPbCl$_3$ they use the value of $0.5605$~nm extrapolated from other experimental data in Ref.~\citenum{Moreira07}. In Ref.~\citenum{Moreira07} one may also find the lattice constant $0.5874$~nm for cubic CsPbBr$_3$. Note that the high-temperature lattice constant of cubic CsPbI$_3$ extrapolated to room-temperature\cite{Trots08} would give $0.6249$~nm.

In Ref.~\citenum{Becker18}, the values 0.6238, 0.5865, and 0.5610~nm are used for I-, Br- and Cl-based cubic perovskites. 
Calculations using the WIEN2k package (version 23.2) with the PBEsol exchange-correlation potential and default high-precision settings (\texttt{-prec 3n}) give respectively 0.6243, 0.5856, and 0.5619~nm. These values are in reasonable agreement with experimental data on the lattice constants of low-symmetry CsPb(I,Br)$_2$ in Ref.~\citenum{Brennan18}.


\putbib[perovskite_QDs]
\end{bibunit}

\end{document}